\documentclass[sigconf]{acmart}

\usepackage{microtype}
\usepackage{algorithm}
\usepackage{graphicx}
\usepackage{algpseudocode}
\usepackage{listings}

\usepackage{lipsum}
\usepackage{soul}
\usepackage{enumitem}
\usepackage{multirow}
\newcommand{\redtext}[1]{\textcolor{red}{#1}}
\setlist[itemize]{leftmargin=20pt} 

\AtBeginDocument{%
  }
    
\copyrightyear{2025}
\acmYear{2025}
\setcopyright{acmcopyright}
\setcctype{by}
\acmConference[MICRO '25]{58th IEEE/ACM International Symposium on Microarchitecture}{October 18--22, 2025}{Seoul, Republic of Korea}
\acmBooktitle{58th IEEE/ACM International Symposium on Microarchitecture (MICRO '25), October 18--22, 2025, Seoul, Republic of Korea}\acmDOI{10.1145/3725843.3756027}
\acmISBN{979-8-4007-1573-0/2025/10}


\begin{document}

\author{Kaiyan Chang}
\affiliation{%
\institution{SKLP, Institute of Computing Technology, Chinese Academy of Sciences}
\institution{University of Chinese Academy of Sciences}
\city{Beijing}
\country{China}
}

\author{Wenlong Zhu}
\affiliation{%
\institution{SKLP, Institute of Computing Technology, Chinese Academy of Sciences}
\institution{University of Chinese Academy of Sciences}
\city{Beijing}
\country{China}
}
\author{Shengwen Liang}
\affiliation{%
\institution{SKLP, Institute of Computing Technology, Chinese Academy of Sciences}
\city{Beijing}
\country{China}
}
\author{Huawei Li}
\affiliation{%
\institution{SKLP, Institute of Computing Technology, Chinese Academy of Sciences}
\institution{University of Chinese Academy of Sciences}
\city{Beijing}
\country{China}
}
\author{Ying Wang}
\affiliation{%
\institution{SKLP, Institute of Computing Technology, Chinese Academy of Sciences}
\city{Beijing}
\country{China}
}
\authornote{Corresponding author.}
\title{LLMulator: Generalizable Cost Modeling for Dataflow Accelerators with
Input-Adaptive Control Flow}




\begin{abstract}
Precise and rapid performance prediction for dataflow-based accelerators is essential for efficient hardware design and design space exploration. However, existing methods often fall short due to limited generalization across hardware architectures, applications, and input-dependent control flows.

Considering the rich program semantic knowledge contained in pre-trained large language models (LLMs), which is used for text and code generation, we propose a progressive numeric modeling paradigm based on pre-trained LLMs. This is an approach to achieve hardware, application, and control flow-sensitive generalization in dataflow accelerator performance prediction. Specifically, to make accurate performance estimates for unseen applications beyond the scope of the training data, we propose a numeric prediction model capable of estimating any performance range. This is achieved by treating the numerical data of the dataflow program as separate tokens and using categorical output for performance values, allowing us to observe confidence at each numerical position. Second, LLMulator supports input-adaptive performance prediction by introducing a reinforcement learning-based dynamic calibration framework, enabling accurate modeling of applications whose control flow varies with input—unlike prior methods that assume fixed execution paths. The cycles prediction error converges to within 11.2\% after several iterations, reducing the error by 9.7\% over static models. Finally, to generalize across diverse hardware architectures and configurations, we introduce a progressive data augmentation framework that systematically generates rich datasets spanning software and hardware variations. This includes multi-level dataflow structures and a structured representation of memory parameters and loop mapping primitives, significantly improving the performance of LLMulator for diverse hardware configurations.

Experiments show that LLMulator achieves an average absolute percentage error (MAPE) of 12.2\% in ASIC design scenarios, reducing errors by 16.7\%-7.8\% compared to existing SOTA methods (TLP, GNNHLS). These results establish LLMulator as a highly accurate and interpretable framework for performance prediction, showcasing new opportunities for applying pre-trained LLMs to hardware modeling.


\end{abstract}



\maketitle
\sloppy
\pagenumbering{arabic} 
\setcounter{page}{1}
\section{Introduction}\label{sec:intro}
\begin{figure}
    \centering
    \includegraphics[width=\linewidth]{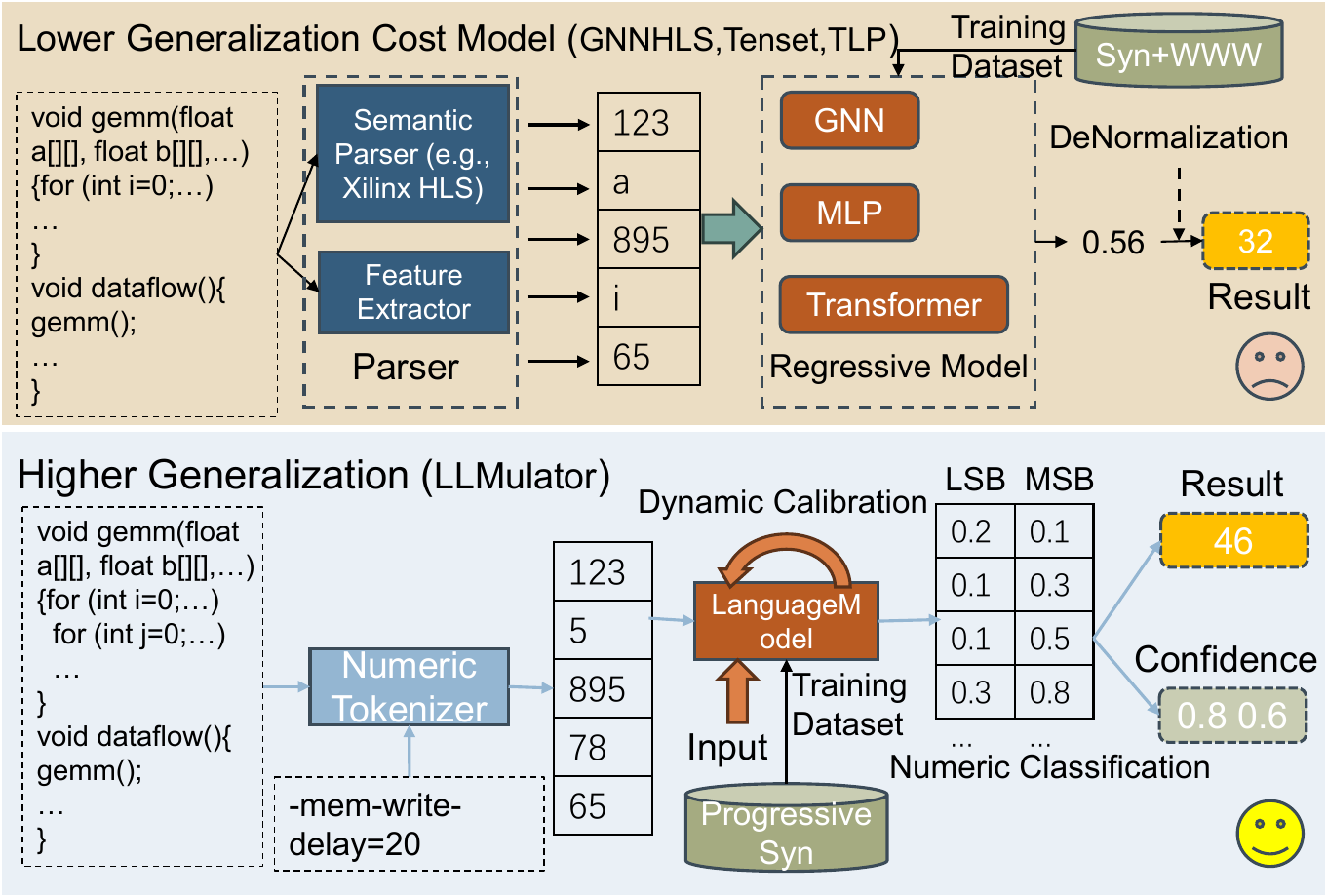}
    \caption{Motivation. LLMulator is an end-to-end cost modeling approach without parser and it produces the result with confidence. $WWW$ denotes the World Wide Web.}
    \label{fig:motivation}
\end{figure}
The rapid evolution of heterogeneous computing architectures has positioned dataflow accelerators (\emph{e.g.,} TPUs, NPUs) ~\cite{neurachip,tpu,moon2021evaluating} as critical infrastructure for key workloads such as deep learning and scientific simulations. Accurate performance prediction for dataflow-based accelerators is essential in early design stage, enabling designers and design space exploration tools to quickly assess and obtain insights into potential design candidates ~\cite{li2022lisa, huang2021cosa, bi2023heron, chen2018learning,xing2022bolt,jung2021deepcuts, hgl, xia2024optimizing, cummins2021programl, wang2024soter, yang2020interstellar, vasilache2018tensor, baghdadi2019tiramisu, zhao2021akg}.

Current performance prediction methods broadly fall into three categories: (1) rule-based simulators, (2) graph neural network (GNN)-based methods, and (3) language model regression techniques, as summarized in Table \ref{tab:simulatorcomp}. 
\textbf{Rule-based simulators} (\emph{e.g.,} Timeloop~\cite{parashar2019timeloop}) rely on manually defined models of hardware behavior. However, their coverage is limited by predefined rules, restricting their ability to generalize to novel or dynamic operators (\emph{e.g.,} dynamic sparse convolutions) and complex control flow scenarios~\cite{yu2018dynamic} (\emph{e.g.,} conditional branches). 
\textbf{Graph neural network} (\textbf{GNN}~\cite{fey2019fast,hamilton2017inductive})-\textbf{based methods} (\emph{e.g.,} GNNHLS ~\cite{wu2022high,cummins2021programl}) compile dataflow programs into graph representations to leverage structural information during inference~\cite{cummins2021programl}. However, they primarily address static graph structures and do not effectively incorporate runtime-sensitive features such as dynamic tensor sizes or loop bounds. Consequently, they yield high prediction errors—up to 27.7\%—in scenarios involving dynamic operator behaviors. 
\textbf{Language model regression methods} (\emph{e.g.,} TLP~\cite{zhai2023tlp}) directly map program text to performance metrics. These data-driven models, which learn from real-world data, offer greater accuracy and flexibility 
 ~\cite{zheng2020flextensor, ansor, baghdadi2021deep, wu2022high,makrani2019xppe,lin2020hl,zhou2019primal,apollo,li2022noception}, which can be represented as a unified workflow in the brown area of Figure \ref{fig:motivation}, consisting of two components: the parser and the regression model. However, their prediction accuracy dramatically deteriorates when encountering large-scale loops or edge cases that exceed the scope of their training data, resulting in errors up to threefold higher in extreme cases.
Collectively, these methods suffer from critical generalization issues across three dimensions: application diversity, input adaptivity, and hardware variability.

\begin{table}[htbp]
\caption{Dataflow Level Cost Modeling Approach Comparison.}
\label{tab:simulatorcomp}
\resizebox{\linewidth}{!}{%
\begin{tabular}{|p{2cm}|p{2cm}|p{2cm}|p{2cm}|}
\hline
\multicolumn{1}{|c|}{\textbf{}} & \multicolumn{1}{c|}{\textbf{Rule-Based}} & \multicolumn{1}{c|}{\textbf{GNN-based}} & \multicolumn{1}{c|}{\textbf{LM Regression}} \\ \hline
Representative   Tools & Timeloop~\cite{parashar2019timeloop}, Maestro~\cite{maestro}, Tenet~\cite{tenet} & GNNHLS~\cite{wu2022high,cummins2021programl} & TLP~\cite{zhai2023tlp} \\ \hline
Technical   Principle & Manually modeled hardware performance predicted through mathematical   formulas & Compiling dataflow programs into   graph structures for inference & Mapping program features to   performance metrics \\ \hline
Applicable   Scenarios & Specific scenarios (\emph{e.g.,} DNNs) & Static dataflow & Fixed operator scheduling \\ \hline
Disadvantages & \begin{tabular}[c]{@{}p{2cm}@{}}1. Cannot generalize to new   operators.\\      2. Reduced edge accuracy due to model generalization\end{tabular} & \begin{tabular}[c]{@{}p{2cm}@{}} 1. Cannot generalize to dynamic   input and dataflow loop extensive applications.\\      2. Reduced edge accuracy due to model generalization\end{tabular} & \begin{tabular}[c]{@{}p{2cm}@{}} 1. Cannot generalize to new   hardware. \\ 2. Cannot generalize to input. \\ 3. Reduced edge accuracy due to regression mode.\end{tabular}\\ \hline
Typical   Error Range & Not explicitly quantified, but   accuracy is limited by manual rules. & Errors in dynamic operator   scenarios up to ~50\%. & Extreme latency errors are three   times higher than those in the central region. \\ \hline
\end{tabular}%
}
\end{table}

\paragraph{Application Generalization.} 
Rule-based simulators (\emph{e.g.,} Timeloop) inherently restrict their applicability to predefined operator structures, often limited to perfectly nested loops without complex control flows (\emph{e.g.,} perfect nested loops without control flow). For example, the ADI application in Polybench cannot be described by Timeloop~\cite{polybench}. Meanwhile, regression-based methods such as TLP are bounded by normalized predictions within training ranges, causing significant errors (MAPE exceeding 40\%) in extreme loop-scale scenarios.  As a result, the cost model often struggles to accurately predict performance for dataflow applications with operators and task graphs that fall outside its training scope.

\paragraph{Input Generalization.} 
Existing cost models  (\emph{e.g.,} Timeloop, TLP) typically ignore input-dependent variations, although inputs often critically influence program performance, particularly through control-flow impacts~\cite{eeckhout2004control, Krogmann2010PerformancePrediction, josipovic2021buffer}. Even slight mispredictions in conditional branch activations can yield substantial deviations in performance estimations. Moreover, incorporating varied inputs together with program and hardware description significantly extends the model input length, highlighting the potential advantage of large-context models such as Large Language Models (LLMs).

\paragraph{Generalization of Hardware Architecture and Mapping Parameters.} 
Hardware parameters—such as loop mapping pragmas and memory delay characteristics—fundamentally shape dataflow performance. Nevertheless, most data-driven models, like GNNHLS~\cite{wu2022high,cummins2021programl} or TLP, are trained on datasets lacking comprehensive hardware configurations. Thus, these models struggle to generalize to unobserved hardware scenarios. These observations motivate the development of a data augmentation framework that incorporates hardware architecture and mapping parameters, aiming to enhance the training of performance prediction models.

To overcome these limitations, we leverage recent advancements in pre-trained LLMs (\emph{e.g.,} LLM for generation~\cite{kao2023flat, devlin2019bert, radford2018improving}, LLM for regression~\cite{vacareanu2024from}). Although LLMs are traditionally applied for text and code generation~\cite{rtlcoder,dataisallyouneed}, their sophisticated semantic understanding provides an underexplored opportunity for numerical performance prediction, with their regression potential demonstrated in~\cite{vacareanu2024from,requeima2024llm}. Inspired by this capability, we introduce \textbf{LLMulator}, the first framework harnessing pre-trained LLMs for generalized, interpretable performance modeling of dataflow accelerators. As illustrated in Figure~\ref{fig:motivation}, LLMulator uniquely incorporates numeric modeling-based static prediction, dynamic prediction-based calibration, and comprehensive dataset augmentation techniques. The contributions of this paper include:

\begin{itemize}
    \item \textbf{Progressive numeric modeling for application generalization.} Our numeric modeling paradigm tokenizes numerical dataflow program elements, enabling performance predictions across previously unseen application scales with significantly reduced edge-value errors. By treating the output numeric values in the dataflow program as distinct tokens and using categorical outputs for performance values, we can observe the confidence at each numeric position. This approach reduces edge value prediction errors by \textit{2.3\%-13.6\%} on the benchmark, outperforming traditional regression methods.
    \item \textbf{Dynamic calibration for input generalization.} To fully leverage input data and enhance prediction accuracy, we introduce a framework for input-driven performance prediction calibration. By tracking real-profile changes under dynamic input variations, we generate preference datasets that are then used to refine the prediction model through reinforcement learning from execution feedback. This method significantly outperforms static modeling techniques by \textit{9.7\%-12.5\%} on the benchmark, reducing discrepancies in dataflow accelerators' static simulations and physical implementations.    
    \item \textbf{Progressive data generation framework for hardware architecture and parameter generalization.} To generalize cost prediction across diverse hardware architectures and parameters, we propose a collaborative data augmentation framework that encompasses both software and hardware to train the LLMulator predictor. This framework first generates multi-level data to represent complex dataflows and then categorizes hardware mappings into memory-related parameters and loop mapping primitives to simplify data permutations and combinations. By profiling intermediate compiler results, this foundational dataset enables the generation of reasoning data, reducing prediction errors from \textit{27.1\%} to \textit{14.2\%}.
\end{itemize}

\section{Background \& Motivation}\label{sec:motiback}

\paragraph{Analysis-based Evaluation Model.} 
As a representative work of this approach, Timeloop \cite{parashar2019timeloop} is a widely used framework for modeling various dataflows on custom accelerators, predicting performance and energy consumption using spatial and temporal notations. Other tools, such as Maestro \cite{maestro, kwon2020maestro} and Tenet \cite{tenet}, present analytical cost models that estimate execution time and energy efficiency by incorporating data-centric directives. Crop ~\cite{crop} decomposes hardware and software features, facilitating the reuse of software components to realize fast and accurate prediction. Soter \cite{wang2024soter} proposes an analytical model to generate a high-quality design space by excluding invalid and inefficient program configurations. Tileflow \cite{zheng2023tileflow} uses a tree-based approach to model dataflow fusion, representing the design space through loop constructs and allowing for detailed performance analysis.

Such methods are categorized as analysis-based evaluation models, which leverage known programmer insights to mathematically model the behavior of spatial accelerators and solve for performance metrics. However, this approach has two main drawbacks. First, the mathematical models are hand-crafted by human and often limited to specific domains, such as deep neural networks (DNNs), and may fail to capture runtime behavior or unstructured dataflows. For example, Timeloop relies on explicitly-specified neural network operator templates, such as \texttt{Hstride} and \texttt{Wstride}, which may not generalize well to other domains. Second, designing such models often requires significant human efforts, such as the complex resource tile-binding analysis used in Tileflow \cite{zheng2023tileflow}. While these models can provide valuable insights, they struggle to account for the full range of variability found in dynamic, real-world execution environments.

\paragraph{Data-driven Evaluation Model.}
Tenset \cite{zheng2021tenset} is a performance prediction dataset specifically designed for tensor operators, associating schedule primitives with their corresponding performance. TVM \cite{chen2018tvm} utilizes this dataset to train multi-layer perceptron (MLP) cost models for individual operators, then optimizing scheduling decisions based on performance data. The availability of datasets highlights the potential of data-driven evaluation methods. HLSGNN~\cite{wu2022high}, for example, employs a graph neural network (GNN) model to estimate performance using high-level synthesis (HLS) intermediate representations. Building on this trend, TLP \cite{zhai2023tlp} introduces a tensor program cost model that treats schedule primitives as sequences of natural language tokens, using a language-based deep learning model to predict the performance of tensor operators based on the specified schedule primitives.
However, predicting the performance of a holistic dataflow program presents additional challenges. Unlike previous methods that rely primarily on input schedule primitives tied to fixed operators (\emph{e.g.,} mapping annotations), holistic program performance prediction must also adapt to comprehensive dataflow program details and hardware characteristics that cannot be covered by pre-fixed schedule primitives. Additionally, the larger representation program space in the graph-based neural network requires more specific data to train, resulting in lower accuracy. 
\begin{figure}[htbp]
    \centering
    \includegraphics[width=\linewidth]{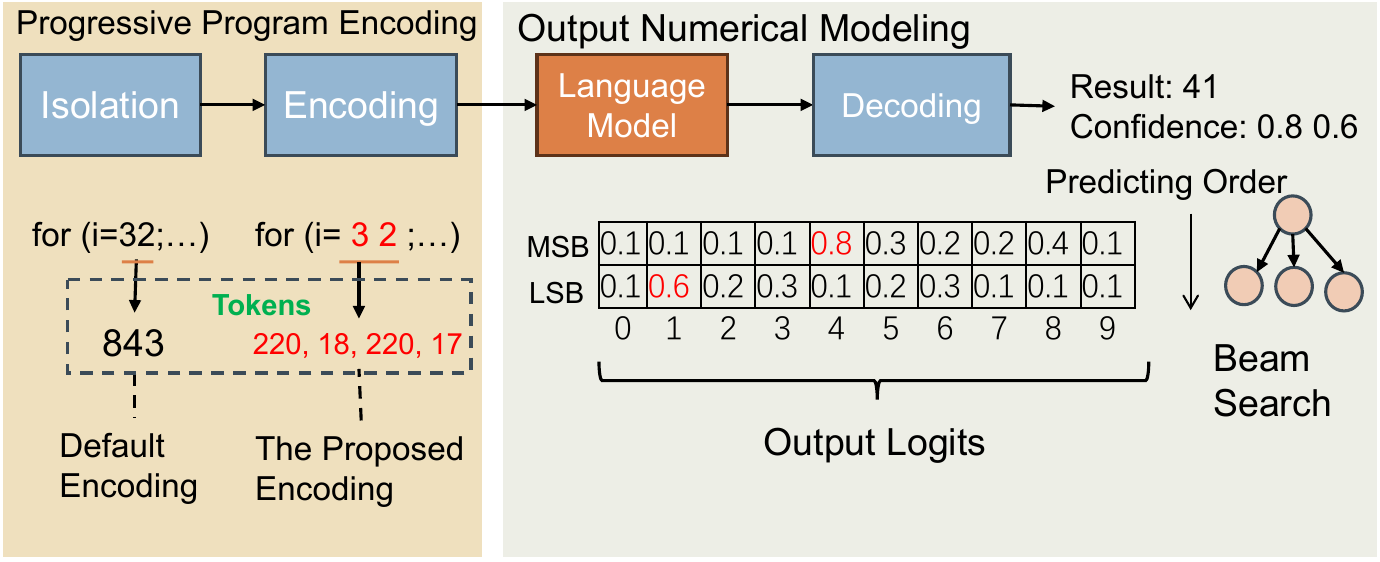}
    \caption{Numeric Modeling. The brown section represents the program encoding, while the green section highlights the numerical modeling strategy for outputs.}
    \label{fig:numericencoding}
\end{figure}

\begin{figure*}[htbp]
    \centering
    \includegraphics[width=.75\linewidth]{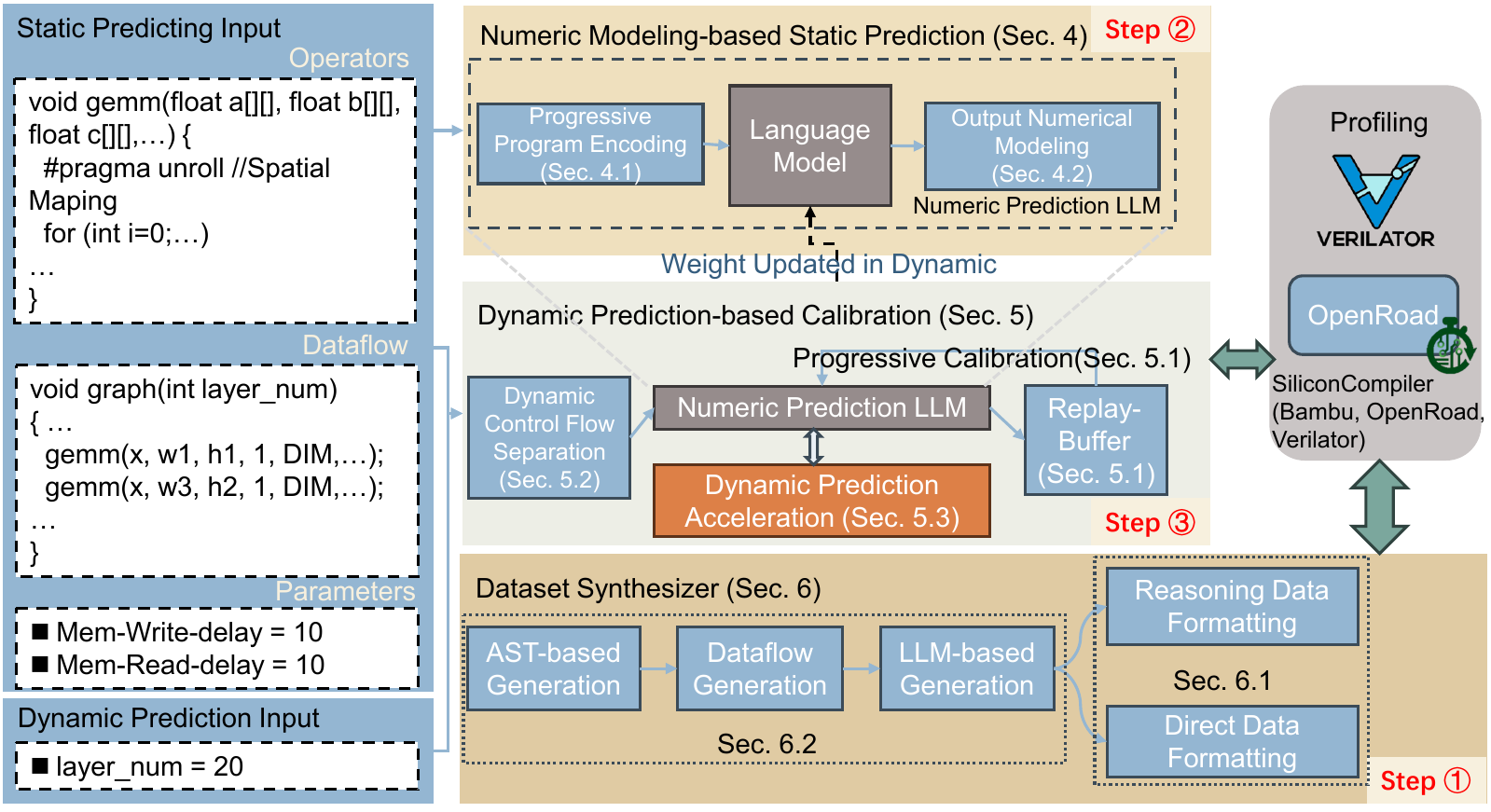}
    \caption{Overview of the LLMulator framework.}
    \label{fig:overview}
\end{figure*}

\paragraph{Challenge 1: Application Generalization}
As highlighted in Section \ref{sec:intro}, existing performance prediction methods commonly struggle with application generalization, limiting their practical use across diverse scenarios. This issue primarily arises from two aspects: numerical precision distortion and semantic information loss.

From the perspective of \textbf{numerical precision distortion}, traditional transformer-based prediction models typically employ regression-based outputs, using normalization techniques such as the Sigmoid function to constrain predictions within a range (0, 1) and employing Mean Squared Error (MSE) as the loss metric. This approach, however, introduces two critical drawbacks:
\begin{itemize}
    \item \textbf{Numerical range compression distortion.} When the target value domain spans a large range (\emph{e.g.,} 0–10000ms), normalization weakens the ability to represent extreme values. For instance, the gradient of the Sigmoid function near its saturation zones (around 0 or 10000ms) approaches zero, making parameter updates during backpropagation difficult.
    \item \textbf{Uneven error distribution.} Experimental results (specific experimental numbers or benchmark results to be added) show that in the edge regions of the target value domain (\emph{e.g.,} <100ms or >9500ms), prediction errors can be more than three times greater than in the central region (\emph{e.g.,} 4500–5500ms), with the highest relative error surpassing 40\%. Moreover, this error is not accompanied by a confidence value for the model user.
\end{itemize}

Moreover, from the viewpoint of \textbf{dataflow program semantic loss}, current language-model-based techniques, such as TLP~\cite{zhai2023tlp}, suffer from substantial information degradation when encoding numerical values within dataflow programs. As illustrated in Figure~\ref{fig:numericencoding}, conventional tokenization methods often encode entire numeric values (\emph{e.g.,} loop boundary values such as 100) as single tokens or irregular splits (\emph{e.g.,} splitting '100' into '10' and '0'), which breaks the semantic coherence of numeric representations. Consequently, the model encounters significant difficulty generalizing predictions to numeric values outside its trained tokenization range. Our evaluations demonstrate that this semantic distortion can lead to prediction accuracy reductions of up to 13.5\%.

To overcome these limitations, we propose a numerical modeling framework that enhances application generalization through the design of a numeric-sensitive output decoder and dataflow input encoder.

\paragraph{Challenge 2: Input Generalization} 
As states in Section \ref{sec:intro}, classical static paradigms fail to generalize to different input variations, and the change of input leads to a decline in prediction accuracy of such models that do not accept program input variation:

\begin{itemize}

    \item \textbf{Control flow sensitivity limitation.} When the program’s control flow heavily depends on input features (\emph{e.g.,} conditional branch probabilities, loop iteration counts), static models fail to cover all possible execution path combinations. As the input distribution shifts, the prediction error accumulates. For example, consider the common sliding window operator in dynamic dataflows: its loop boundaries are dynamically determined by the input tensor size $[H, W]$. While static models may only train on cases where \(H, W \leq 256\), deploying an input where \(H = 1024\) leads to prediction errors surging to 60.2\%, compared to a 10.0\% error on the training set.
    \item \textbf{No support of runtime parameter tracking.} Existing models generally do not update their internal knowledge to reflect runtime design parameters like loop iterations, continuing to rely exclusively on static, pre-trained profile data. This approach prevents capturing performance variations introduced by dynamic changes in dataflow parameters (\emph{e.g.,} during iterative design space exploration~\cite{zheng2020flextensor,zheng2023tileflow}). For instance, when adjusting loop unroll parameters in dataflow designs, static models fail to integrate real-time tuning data, causing prediction inaccuracies.
    
\end{itemize}

These challenges arise from the inherent mismatch between static performance modeling paradigms and dynamic program execution environments. To address this, we propose a dynamic predicting calibration (DPC), which constructs a real-time updating learning system using live profile feedback, fully utilizing input variations sensitive to control flow and mapping changes (detailed in Section \ref{sec:dynamicmethod}).

\paragraph{Challenge 3: Generalization of Hardware Architecture and Mapping Parameters} 
As mentioned in Section \ref{sec:intro}, existing prediction methods often neglect specific hardware design constraints during dataset generation, causing significant domain prediction distribution biases. For instance, synthesized datasets such as those used in GNNHLS typically feature an average loop nesting depth of only 1 layers, whereas real hardware implementations commonly have an average depth of  >3 layers. Additionally, synthesized data often include approximately 10\% non-array operations, whereas practical implementations predominantly employ array-based operations (over 90\%). These discrepancies between synthetic and real hardware architectures limit the predictive model's ability to generalize across hardware.

Moreover, traditional syntax-driven data augmentation lacks intermediate compilation-level reasoning details, relying solely on end-to-end profile outputs. When the predictive model processes dataflow inputs (computation graph and operator), the absence of intermediate features such as module instantiation counts or multiplexer numbers significantly hampers accuracy. For example, directly predicting physical-level information like area results in mean absolute percentage errors (MAPE) as high as 27.1\%.

To tackle these issues, we propose a progressive generation framework. Our approach sequentially applies Abstract Syntax Tree (AST)-based generation, dataflow-specific generation, and large language model (LLM)-based diverse data generation. Additionally, we classify hardware mappings into memory-related parameters and loop mapping primitives, enabling hierarchical generation of foundational data closely aligned with realistic hardware and dataflow distributions.

\section{LLMulator Overview}\label{sec:overview}

LLMulator is a comprehensive, generalizable performance prediction framework tailored for dataflow-based hardware-software co-design scenarios. Its hierarchical modeling strategy systematically addresses the limitations identified in traditional static approaches, as illustrated in Figure \ref{fig:overview}.  The framework receives input as a quadruple $\{G, Op, \allowbreak Params, data\}$, where $G$ represents the dataflow graph program, ${Op}$ denotes specific operator implementations, $Params$ encompasses hardware configuration, and $data$ (represented as "\texttt{[variable name]}$\allowbreak =$\texttt{[value]}") specifies runtime inputs. LLMulator outputs a multidimensional performance metric vector comprising $\langle Power, \allowbreak Area, Flip-Flop, Cycles \rangle$. The core of LLMulator comprises three interconnected stages: 

\begin{itemize}
\item \textbf{Dataset synthesizer.} This component systematically generates comprehensive training data, capturing various hardware-software interaction scenarios. It employs a progressive generation mechanism, integrating Abstract Syntax Tree (AST)-based generation, dataflow-specific generation, and Large Language Model (LLM)-enhanced diversity in data synthesis, producing datasets aligned closely with realistic hardware implementations. 
\item \textbf{Numeric modeling-based static prediction.} Leveraging the synthesized dataset, the static prediction mode estimates hardware-level metrics such as power consumption, chip area, and achievable clock periods using compile-time information. This prediction stage addresses application-level and hardware-level generalization by employing a specialized numeric-sensitive decoder and a semantic-aware encoding of numeric values within dataflow programs.

\item \textbf{Dynamic prediction-based calibration.} Recognizing the inherent variability introduced by dynamic inputs, this stage employs real-time profiling feedback to refine cycle-count predictions. It continuously adapts and improves prediction accuracy through a reinforcement learning-based calibration mechanism, effectively managing variations in control flow, loop bounds, and runtime parameters.
Together, these stages ensure accurate, robust, and adaptable performance predictions, enabling precise performance modeling across diverse dataflow scenarios.\end{itemize}

\section{Numeric Modeling-based Static Prediction}  \label{sec:numericalmodeling}
Although LLMs demonstrate exceptional capabilities in semantic program understanding, their native implementations often exhibit limited numeric comprehension within program contexts, as detailed in Section \ref{sec:motiback}. To address these numeric prediction challenges, this section introduces a progressive numerical prediction framework, illustrated in Figure \ref{fig:numericencoding}.

\subsection{Progressive Program Encoding}  
To address the dataflow program encoding distortions in application generalization discussed in Section \ref{sec:motiback}, we observe that pre-tokenization can significantly improve LLM performance~\cite{wu2024pre,dagan2024getting}, then design a progressive tokenizer architecture with dual-phase processing to preserve numerical semantics in LLM-based dataflow performance prediction:

\begin{itemize}  
    \item \textbf{Symbol isolation phase.} We insert protective half-character spaces between numerical values based on programming language syntax. For example, "-128" becomes " - 128", ensuring independent encoding of symbols and numbers while preventing structural disruption.
    \item \textbf{Encoding phase.} The standard tokenizer is applied to the isolated text stream, achieving a linear correlation between numerical token count and digit length (length $n$ → $n$ Tokens), aligning with human cognitive patterns.  
\end{itemize}  

Section \ref{sec:expeablation} demonstrates that progressive program encoding reduces numerical recognition errors from 23.7\% to 10.2\% on power prediction.

\subsection{Output Numerical Modeling}\label{sec:outputnumericmodeling}  
To mitigate numerical range compression distortion in application generalization, as outlined in Section \ref{sec:motiback}, we propose a classification-based numeric prediction approach, designed to enhance edge precision and embed explicit confidence indicators in performance predictions as shown in Figure ~\ref{fig:numericencoding}. Key innovations include:

\begin{itemize}  
    \item \textbf{Decoupled numerical modeling.} We decompose continuous numerical predictions into progressive digit-wise classification tasks, moving hierarchically from high-order to low-order digits. Each digit prediction is modeled as an independent classification problem with classes ranging from 0–9. For example, predicting a value of 655 involves first identifying the hundreds digit "$6$" from a constrained candidate set $\{0,1,...,9\}$, subsequently refining predictions for tens and units digits "$5$". This hierarchical prediction aligns naturally with animals decimal cognition\footnote{In book \textit{The Number Sense}~\cite{dehaene2011number} page 27: "The animals picked out the larger digit with a much higher success rate than chance alone would have predicted."}.

    \item \textbf{Error control mechanism.} 
    Our hierarchical digit-wise classification employs a beam search strategy to manage errors effectively. When higher-order digit predictions are erroneous, subsequent lower-order digits can partially rectify these inaccuracies (\emph{e.g.,} adjusting "7XX" to "655" using beam search). Each digit classification generates logits indicating confidence, allowing the model to systematically choose digit candidates with maximum confidence and rectify initial prediction errors.

    \item \textbf{Interpretability enhancement.} By presenting the probability distributions (logits) for digit predictions, the method transparently communicates model confidence. For instance, a bimodal distribution (\emph{e.g.,} "4:0.8, 1:0.6" in Figure \ref{fig:numericencoding}) indicates uncertainty in boundary value predictions, enhancing interpretability for users.

\end{itemize}

Furthermore, we observe that the choice of base encoding (denoted as \( D \)) in progressive numerical modeling fundamentally represents a trade-off between encoding length and classification complexity. For target value \( N \), its encoding length (temporal efficiency) under base \( D \) is \( L = \lceil \log_D N \rceil \), while per-digit classification complexity (spatial efficiency) is \( D \). This yields:  

\begin{itemize}  
    \item \textbf{Temporal efficiency.} Larger \( D \) logarithmically reduces \( L \) (\emph{e.g.,} \( N=128 \): \( D=10 \Rightarrow L=3 \), \( D=2 \Rightarrow L=7 \));  
    \item \textbf{Spatial efficiency.} Per-digit classification complexity (logit dimension) grows linearly with \( D \), increasing model parameters and computational costs.  
\end{itemize}  

For \( N=128 \):  
\begin{itemize}  
    \item \textbf{Decimal (\( D=10 \)).} Encoding [1, 2, 8], logit dimension 10, requiring 3 prediction steps;  
    \item \textbf{Binary (\( D=2 \)).} Encoding [1, 0, 0, 0, 0, 0, 0], logit dimension 2 but requiring 7 steps, introducing long-range dependency issues.  
\end{itemize}  

Kolmogorov complexity theory~\cite{li2008introduction} further explains base selection trade-offs: smaller \( D \) aligns better with the inductive bias of text sequence models but leads to long-range dependencies, while larger \( D \) facilitates parallelism but increases per-step complexity. This helps explain why classical transformer-based regression cost prediction models~\cite{zhai2023tlp} (with an infinite base) achieve lower accuracy compared to our progressive numerical encoding approach. 

Therefore, we utilize the categorical cross-entropy loss function, as specified in Equation \ref{equ:cce}, rather than the mean squared error  typically employed in regression models~\cite{zhai2023tlp,wu2022high}. This adjustment enables the prediction of the $j$-th digit, where 
$y_i$ represents the true one-hot encoded value for the $i$th candidate logit, and $\hat{y_i}$ denotes the predicted $i$th logit probability.

\begin{equation}\label{equ:cce}
    L_j=-\sum_{i=1}^D y_ilog(\hat{y_i}), 0\le j<L
\end{equation}


\section{Dynamic Prediction-based Calibration}\label{sec:dynamicmethod}
While the basic static cost prediction was discussed in the previous section, the cost model still lacks adaptability to different inputs. This section presents solutions to the input generalization challenge, including a dynamic prediction calibration mechanism and a dynamic control flow separation mechanism to analyze and optimize the solution generated by the static stage.
\subsection{Dynamic  Predicting Calibration} \label{sec:dynamicpredcali}

\begin{figure}[htbp]
    \centering
    \includegraphics[width=.8\linewidth]{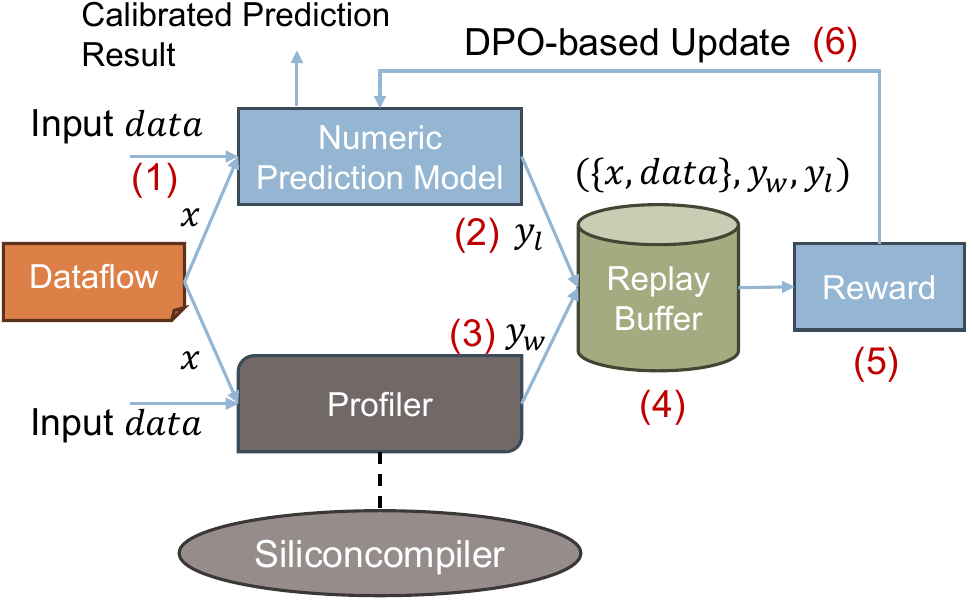}
    \caption{Dynamic Predicting Calibration Flow via Direct Preference Optimization.}

    \label{fig:dynamiccalibration}
\end{figure}

Our dynamic prediction calibration mechanism transforms performance prediction into an adaptive online learning framework, as illustrated in Figure \ref{fig:dynamiccalibration}. Here, the model continuously interacts with the simulation environment, enabling real-time performance refinement. Specifically, the predictive model  outputs predictions  for input dataflow configurations, and the environment returns actual performance  from execution simulations (\emph{e.g.,} via verilator~\cite{verilator}). The calibration leverages the direct preference optimization (DPO)~\cite{rafailov2023direct} concept from reinforcement learning (\emph{i.e.,} RLHF~\cite{lee2023rlaif}). Unlike standard RL, which often needs a separate "reward model", DPO directly optimizes the prediction model using preference data. We adapt it to model real-profile feedback through the following iterative (Figure ~\ref{fig:dynamiccalibration}):

\begin{enumerate}
  \item \textbf{Input selection:} 
    The numeric prediction model takes as input the dataflow design \(x\) (\emph{i.e.,} the dataflow text $\{G,Op,Params\}$ ) together with dynamic input \(data\) (\emph{i.e.,} the application dataflow input). This forms the state \(\{x, \text{data}\}\).
  \item \textbf{Prediction:} 
    The model \(f_{\theta}\) produces a predicted performance \(y_l = \hat y = f_{\theta}(x, \text{data})\), where $\theta$ represents model weights.
  \item \textbf{Profiler feedback:} 
    The environment (\emph{e.g.,} Siliconcompiler) returns the “ground-truth” performance \(y_w\) for the same \(\{x, \text{data}\}\).
  \item \textbf{Dynamic preference data pairs:} 
    Construct a preference triplet \(\bigl(\{x,\text{data}\}, y_w, y_l\bigr)\). 
    For example, if the model predicts \(y_l=1000\) cycles but the actual is \(y_w=950\) cycles, the pair \((950, 1000)\) indicates that 950 is “better.” DPO uses this to adjust \(\theta\) so future predictions for similar inputs move closer to \(y_w\).
  \item \textbf{Real-profile reward function:} 
    The calibration defines a direct reward function (see Equation~\ref{equ:dppc}) that quantifies how much \(y_w\) is preferred over \(y_l\). A reference policy \(\pi_{\mathrm{ref}}\) (\emph{i.e.,} the prediction model pre-trained in the static stage) provides a baseline. The reward ensures alignment of predictions with the real performance distribution, with parameters \(\beta\) controlling sensitivity and the Sigmoid function \(\sigma\) for normalization.
  \item \textbf{DPO-based update:} 
    By leveraging the preference pair \((y_w, y_l)\) and the real-profile reward in Equation~\ref{equ:dppc}, we directly update the model parameters \(\theta\) via gradient descent to better align predictions with profiling results for similar states.
\end{enumerate}

\begin{equation}\label{equ:dppc}
    \resizebox{\linewidth}{!}{%
    $\mathcal{R}(\theta) = \mathbb{E} \left[ \log \sigma \left( \beta \left( \log \frac{\pi_\theta(y_w|\{x,data\})}{\pi_{\text{ref}}(y_w|\{x,data\})} - \log \frac{\pi_\theta(y_l|\{x,data\})}{\pi_{\text{ref}}(y_l|\{x,data\})} \right) \right) \right]$
    }
    \end{equation}

\subsubsection*{Replay-cost-buffer design.} To reduce overfitting to specific profiles, we introduce a configurable replay buffer that stores preference triplets \( (\{x,data\}, y_w, y_l) \) within a sliding window. This buffer supports minibatch replay, allowing efficient reuse of recent data and adapting to varying input distributions effectively. A buffer size of 1 enables immediate real-time updates, suitable for stable environments.

By continuously updating parameters based on real-time feedback, this mechanism significantly enhances model adaptability and accuracy, particularly as input distributions evolve.

\subsection{ Dynamic Control Flow Separation}\label{sec:dynamicsepa}
\begin{figure}[htbp]
    \centering
    \includegraphics[width=.65\linewidth]{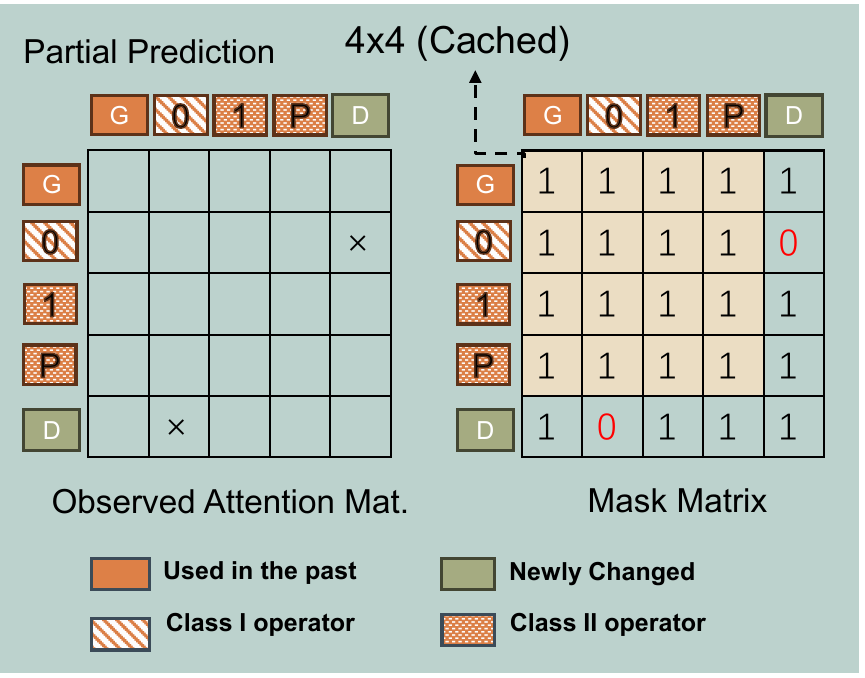}
    \caption{ Dynamic Control Flow Separation: Eliminating redundant operator predictions. G denotes a dataflow graph. 0 denotes $Op_0$. 1 denotes $Op_1$. P denotes parameters. D denotes input data. }
    \label{fig:controlflowsepa}
\end{figure}

To mitigate increased prediction latency from dynamic prediction-based calibration in long dataflow implementations, we introduce dynamic control flow separation. This method enhances efficiency by categorizing prediction targets based on their input dependencies, allowing for separate processing of different input components:

\begin{itemize}
    \item \textbf{Static predictable parameters.} Metrics determinable at compile-time (\emph{e.g.,} area, static power) utilize a static prediction model. The model input vectors, $\{G, Op_0, \allowbreak Op_1, Params\}$, where $Params$ represents hardware mapping and parameters, exclude runtime data from programs.

    \item \textbf{Dynamic dependent parameters.} Metrics influenced by runtime inputs (\emph{e.g.,} cycles) require extended input vectors, $\{G, Op_0, Op_1, \allowbreak Params, data\}$, incorporating runtime data explicitly (scalars separated by commas), where $data$ represents the input data.
\end{itemize}

Predicting all parameters dynamically causes redundant computations. To address this, we categorize operators based on static dataflow analysis: \underline{Class I} operators have input-independent control flow (\emph{e.g.,} matrix transposition), while \underline{Class II} operators' control flow depends on input data (\emph{e.g.,} sorting algorithms). This distinction reveals redundancy in standard attention, as Class I operator predictions (\emph{e.g.,} $Op_0$, independent of data, as shown in Figure \ref{fig:controlflowsepa}) are still associated with dynamic data. To mitigate this, we use pre-analysis metadata to construct a mask matrix, concealing the interaction between Class I operators and dynamic data (set the mask to zero).

\subsection{Dynamic Prediction Acceleration}\label{sec:dynamicaccel}

\begin{figure*}[htbp]
  \centering
  \begin{minipage}[t]{0.28\linewidth} 
    \centering
    \includegraphics[width=\linewidth]{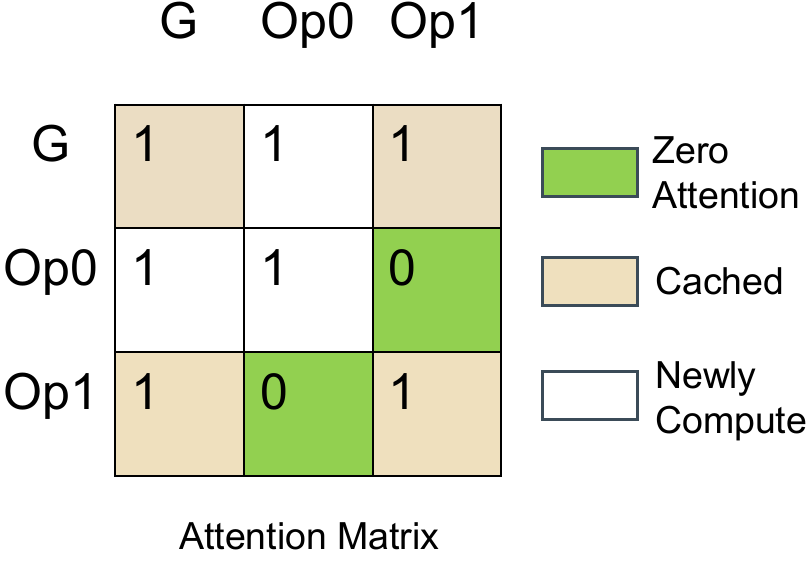}
    \caption{Progressive operator cache with masking with $Op_0\times Op_1$.}
    \label{fig:fastmodelingattn}
  \end{minipage}
  \begin{minipage}[t]{0.68\linewidth} 
    \centering
    \includegraphics[width=\linewidth]{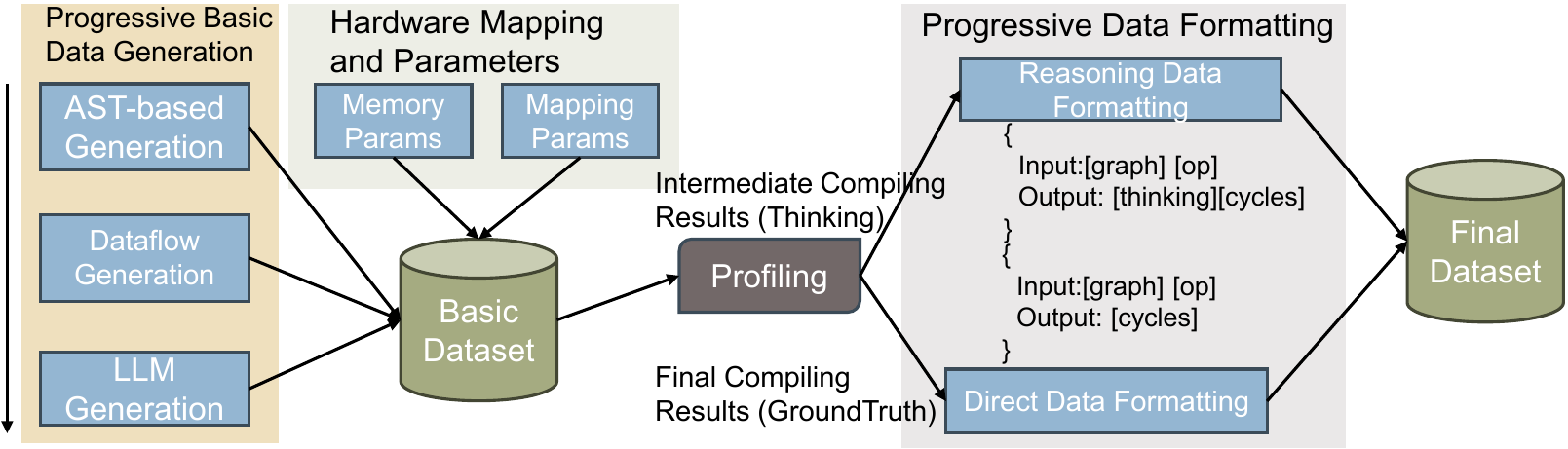}
    \caption{Dataset Synthesizer.}
    \label{fig:datagenerator}
  \end{minipage}
\end{figure*}

\subsubsection*{Acceleration Overview} Adjusting runtime parameters of the workloads (\emph{e.g.,} adjusting the stride of a convolution operator or changing the activation function) typically forces re-evaluation of the entire dataflow graph, introducing substantial computational redundancy. Inspired by partial attention mechanisms~\cite{beltagy2020longformer}, we propose a dynamic prediction acceleration approach:
\begin{itemize}
\item \textbf{Precomputation and caching.} Attention matrices unrelated to dynamic changing operators are cached, allowing for rapid updates of only relevant sub-matrices.
\item \textbf{Selective attention masking.} Mask matrices set the interrelations of unrelated operators to 0, preventing their redundant computation and minimizing unnecessary recalculations, as illustrated in Figure \ref{fig:fastmodelingattn}.
\end{itemize}

This selective recalculation significantly accelerates performance predictions, especially beneficial during iterative design optimizations cost modeling.

\subsubsection*{Acceleration Case Study} Consider the dataflow  $\{G, Op_0, Op_1\}$, where $Op_0$ is the changed operator need to re-evaluate. Because the cost of the two operators are independent (\emph{i.e.,} the two operators' costs do not rely on each other), this forms a decoupled attention pattern. Specifically, as illustrated in Figure \ref{fig:fastmodelingattn},  there are nine regions in the attention matrix. Two types of regions can be precomputed: 

\begin{itemize}
    \item The four corner regions, which do not involve intermediate changing operators, can be computed in earlier stages and thus precomputed.
    \item The two regions associated with $Op_0$ and $Op_1$, which are unrelated, can be pre-initialized to zero.
\end{itemize}

\section{Dataset Synthesizer}\label{sec:datasetgeneration}

To enhance LLMulator's generalization capabilities across diverse hardware architectures and mapping parameters, we introduce a comprehensive data augmentation framework. This framework synthesizes progressive datasets, capturing essential variations and characteristics necessary for robust model training, as illustrated in Figure \ref{fig:datagenerator}.
\subsection{Progressive Basic Data Generation}

The dataset synthesizer follows the progressive construction principle of "general first, then specific," a strategy proven effective in curriculum learning~\cite{bengio2009curriculum} and domain transfer learning~\cite{raffel2020exploring}. We adopt a hierarchical data generation approach, gradually transitioning from general to more specific data representations:

\begin{itemize}

\item \textbf{AST-based generation.} Utilizing an Abstract Syntax Tree (AST)-based generator (\emph{e.g.,} \texttt{ldrgen}), we initially generate basic, syntactically correct dataflow program structures. This ensures fundamental correctness in variable scope and control-flow constructs.
\item \textbf{Dataflow-specific generation.}
\begin{sloppy}
We further specialize dataflow program generation by targeting loops and array operations. Specifically, the dataflow generation process includes a graph generator that randomly changes operator parameters and their order, and an operator generator that creates \texttt{for}-loops. We adapt the tree-based loop cost modeling approach~\cite{zheng2023tileflow} to a tree-based generation method. The operator generator models operators as a loop tree, mutating both the loop order and the step sizes within a specified range, specifically targeting dataflow patterns relevant to hardware mappings. Additionally, when control flow is dependent on input, we iterate over the scalars within the input range. The range is set by applying random variations of -50\% and +50\% to the input scalars. 
\end{sloppy}

\item \textbf{LLM-based generation.} To diversify the dataset beyond template limitations, we leverage LLM-based self-augmentation, which obtains dataflow variant by changing the basic dataflow with prompting. This enables the flexible restructuring and variation of typical dataflow programs, ensuring a broad coverage of realistic scenarios.
\end{itemize} 

The progressive data synthesizer first generates code as the basic dataset, as shown on the left side of Figure \ref{fig:datagenerator}. It then uses the progressive data formatter to inductively generate the final dataset for training (Section \ref{sec:datagenprogressive}), depicted on the right side of Figure \ref{fig:datagenerator}.

\subsection{Progressive Data Formatting}\label{sec:datagenprogressive}
Once the foundational dataset is generated, it is enhanced through progressive data formatting to incorporate profiling results and intermediate reasoning information necessary for model training:

\subsubsection*{Reasoning Data Formatting.} Intermediate representations (IR), such as RTL and netlists, provide valuable reasoning information. Inspired by the reasoning chain concept proposed in \cite{wei2022chain, luo2024improve, guo2025deepseek}, explicitly constructing the reasoning path using the \texttt{<think>} tag can enhance the model's accuracy. However, directly incorporating the entire IR into the reasoning path may exceed context-length limitations. To address this challenge, we propose a progressive encapsulation strategy:

\begin{figure*}[!htbp] 
    \centering 
    \begin{minipage}[t]{0.3\linewidth} 
        \centering 
            \centering
            \setlength{\fboxsep}{1pt} 
            \setlength{\fboxrule}{0.5pt} 
            \fbox{
                \includegraphics[width=\linewidth]{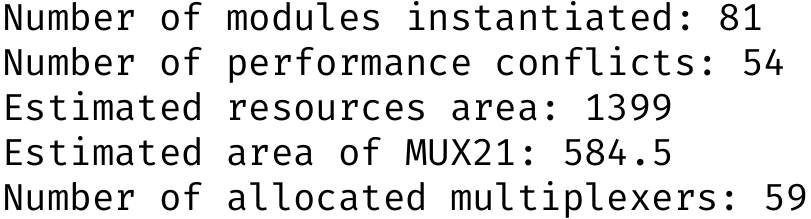}
            }%
            \caption{Thinking Format.} 
            \label{fig:thinkingdata} 
    \end{minipage}
    \hfill 
    \begin{minipage}[t]{0.3\linewidth} 
            \centering
            \setlength{\fboxsep}{1pt} 
            \setlength{\fboxrule}{0.5pt} 
            \fbox{
                \includegraphics[width=\linewidth]{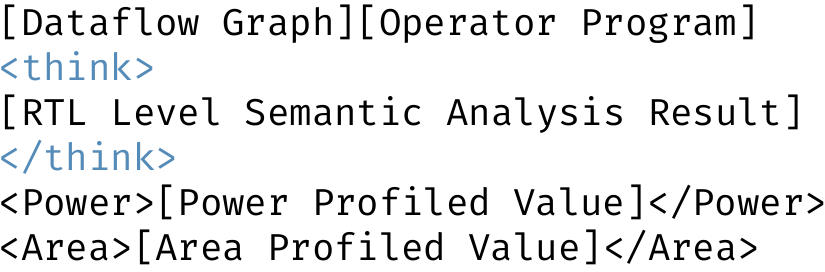}%
            }%
            \caption{Reasoning Data Format.}
            \label{fig:sepadata}
    \end{minipage}
    \hfill 
    \begin{minipage}[t]{0.35\linewidth} 
    \centering
  \setlength{\fboxsep}{1pt}  
  \setlength{\fboxrule}{0.5pt} 
  \fbox{%
      \centering
      \includegraphics[width=\linewidth]{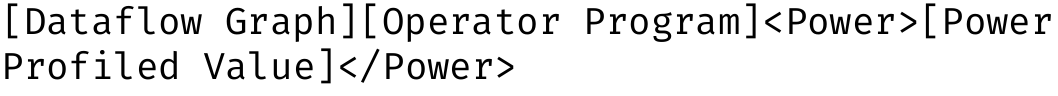}
  }
  \caption{Direct Data Format.}
 \label{fig:unidata}
 \end{minipage}
\end{figure*}

\begin{itemize}
\item \textbf{Intermediate feature extraction.} We utilize SiliconCompiler~\cite{siliconcompiler2022} to extract critical RTL-level features (\emph{e.g.,} module counts, multiplexer numbers), significantly reducing IR size, as depicted in Figure \ref{fig:thinkingdata}.
\item \textbf{Encapsulated reasoning fragments.} Extracted features are encapsulated using a \texttt{<think>} tag, explicitly constructing reasoning paths to enhance model interpretability and accuracy, as illustrated in Figure \ref{fig:sepadata}.

\item \textbf{Dataset construction.} The final dataset comprises the original dataflow program \(P\), the encapsulated reasoning fragments \(R\), and the performance prediction target \(C\), structured as \([P, R, C]\).
\end{itemize}

\subsubsection*{Direct Data Formatting.} For scenarios prioritizing efficiency, we also introduce a direct data format, as depicted in Figure \ref{fig:unidata}, where the input computational graph and operator program directly map to prediction targets without intermediate reasoning steps. This simpler format accelerates dataset creation and suits end-to-end prediction tasks.

\subsection{Hardware Mapping and Parameters}

We systematically introduce critical hardware mapping parameters into our datasets, encompassing both memory-related configurations and loop-mapping primitives:

\begin{itemize}
\item \textbf{Memory-related parameters.} Explicitly configured through HLS tools (\emph{e.g.,} Bambu HLS) with parameters such as memory delays (\emph{e.g.,} \texttt{-mem-delay-read=20}). We include varied delay scenarios (10, 5, and 2 cycles) to cover diverse hardware configurations.

\item \textbf{Loop-mapping primitives.} We employ code-level annotations (\texttt{pragma}) for spatial and parallel mappings. Specifically, \texttt{\#pragma clang loop unroll(full)} and \texttt{\#pragma omp parallel for} directives are applied systematically, capturing over 90\% of valid hardware mapping scenarios in benchmark evaluations.
\end{itemize}

Through this comprehensive progressive data augmentation approach, LLMulator achieves robust and generalized performance prediction capabilities across diverse hardware configurations.

\begin{figure}[h]
  
\end{figure}

\section{Evaluation}\label{sec:evaluation}
\subsection{Evaluation Setup}\label{sec:evalsetup}

\subsubsection*{Profile Data Environment} 
We build our performance profiling pipeline using open-source EDA tools accessible via Python. The static prediction stage relies on the SiliconCompiler framework with Bambu~\cite{bambu2021} as the HLS frontend, which compiles C-based dataflow graphs and operators into Verilog. For ASIC flows, OpenROAD~\cite{ajayi2019openroad} performs physical synthesis and extracts static metrics such as area, longest path delay, and power. Dynamic metrics (e.g., cycle counts) are obtained using Verilator, with runtime inputs provided via XML and compiled through Bambu. We use the SkyWater130nm~\cite{skywater_pdk} process library and disable BRAM allocation to ensure general-purpose datapath evaluation. Our evaluation spans both static metrics (area, power, flip-flop count) and dynamic runtime (cycle count).

\subsubsection*{Model Training Environment} 
The LLMulator model training is conducted on 8 NVIDIA A100 80GB GPUs with an Intel Xeon 8358P CPU. We use LLaMA-3.2-1B~\cite{touvron2023llama} as the base model, implemented via HuggingFace Transformers for SFT and inference.  The TRL~\cite{vonwerra2022trl} library is employed for DPO-based dynamic prediction calibration. For static data generation, we adopt \texttt{ldrgen}~\cite{barany2017liveness}, and Frama-C~\cite{kirchner2015frama} is used for static/dynamic control flow analysis. To mitigate catastrophic forgetting~\cite{kirkpatrick2017overcoming}, we apply LoRA~\cite{hu2022lora} instead of full fine-tuning, enhancing generalization from early AST-based examples. The training spans 5 epochs using AdamW as the optimizer.

To ensure fairness, all baseline models are trained on the same dataset, composed of approximately 30\% AST-based, 50\% dataflow-specific, and 20\% LLM-generated programs. We additionally include data from the dataset used in GNNHLS~\cite{wu2022high}.

\subsubsection*{Benchmark and Measurement Standards} To ensure consistency with previous related work, we select Polybench~\cite{karimov2019polybench} as a standard benchmark and use to it validate LLMulator against baselines in  prior studies. For testing the performance of our framework on modern applications, we choose additional benchmarks in Table ~\ref{tab:benchmarkanalysis} that span two mainstream categories: image processing tasks (1-9) and natural language processing tasks (10-14). The column "All Len" represents the character number of the target workload fed to LLMulator, including dataflow graph and operators. "Graph Len" denotes the number of characters to represent the dataflow graph program, while "Op Len" denotes that of operators in a program. "Op Num" denotes the number of operators in the dataflow graph.  "Dyn. Num" denotes the number of optional dynamic control flow-related parameters in the program, which will change the online execution flow of program. For example, in \texttt{for i=0:N} and $N$ is a function input parameter, then the parameter $N$ is a dynamic number.  To test scenarios where control flow is dependent on input, we modify the input of certain operators during execution: (1) operators with image input data modify the input image size, and (2) operators with text input data modify the input text size.  We use the final logit as the confidence value for the predicted result due to its relevance in causal inference. Model inference temperature is set to $1.0$.

\begin{table}[htbp]
\caption{Benchmark Analysis. }
\label{tab:benchmarkanalysis}
\setlength{\tabcolsep}{0.5pt}
\resizebox{\linewidth}{!}{%
\begin{tabular}{|c|c|c|c|c|c|}
\hline
\textbf{Workloads} & \textbf{All Len} & \textbf{Graph Len} & \textbf{Op Num} & \textbf{Dyn. Num} & \textbf{Op Len} \\ \hline
1-Image   Norm+CNN~\cite{ioffe2015batch,lecun1998gradient} & 7002 & 1483 & 8 & 2 & 5519 \\ \hline
2-RB+DSC~\cite{he2016deep,chollet2017xception} & 4742 & 1700 & 6 & 3 & 3042 \\ \hline
3-SPP+Fusion~\cite{he2015spatial,xu2020u2fusion} & 5754 & 1585 & 8 & 2 & 4169 \\ \hline
4-CBAMAttention~\cite{woo2018cbam} & 6822 & 1273 & 12 &52 & 5549 \\ \hline
5-Anchor+RolAlign~\cite{girshick2015fast,he2017mask} & 4711 & 1418 & 5 & 4 & 3293 \\ \hline
6-Gan+SuperResolu.~\cite{goodfellow2014generative,ledig2017photo} & 7520 & 1023 & 13 & 2 & 6497 \\ \hline
7-Dens+SkipConn.~\cite{iandola2014densenet,he2016deep} & 9622 & 753 & 8 & 3 & 8869 \\ \hline
8-DilatedConv+Aggre.~\cite{yu2015multi} & 4839 & 151 & 6 &  2 & 4688 \\ \hline
9-BEVFormer~\cite{li2024bevformer} & 3997 & 1356 & 5 &  2 & 2641 \\ \hline \hline
10-Bert-base~\cite{devlin2019bert} & 7196 & 1278 & 12 &2  & 5918 \\ \hline
11-Albert~\cite{lan2019albert} & 7443 & 1244 & 13 & 4 & 6199 \\ \hline
12-T5-base~\cite{raffel2020exploring} & 13345 & 1314 & 21 & 1 & 12031 \\ \hline
13-Roberta~\cite{liu2019roberta} & 7290 & 994 & 10 & 2 & 6296 \\ \hline
14-LLaMA~\cite{touvron2023llama} & 6726 & 2099 & 8 & 1 & 4627 \\ \hline
\end{tabular}%
}
\end{table}

\subsubsection*{Metrics} 
We evaluate model accuracy using Mean Absolute Percentage Error (MAPE) and Mean Squared Error (MSE), with pass@5 sampling to reduce randomness. Inference time (seconds) is reported for runtime comparison.

\subsubsection*{Baselines} We select the following state-of-the-art dataflow cost modeling methods as baselines.
\begin{itemize}
\item \textbf{TLP \cite{zhai2023tlp}.} A state-of-the-art language model-based method for predicting dataflow operator performance. Unlike general LLMs, TLP employs a direct regression model that outputs fixed normalized performance values and does not use pre-trained model, which is not a decoder-only structure.

\item \textbf{GNNHLS \cite{wu2022high,cummins2021programl}.} A state-of-the-art framework that converts HLS programs into graphs for cost prediction using graph neural networks (GNNs).

\item \textbf{Tenset-MLP \cite{zheng2021tenset}.} This baseline method employs a multilayer perceptron (MLP)-based cost model to predict operator performance by transforming input tensors into handcrafted features through pre-profiling. It captures limited input variability by extracting coarse-grained indicators such as loop bounds or tensor dimensions. However, it treats all inputs with the same loop range or shape as equivalent, ignoring finer-grained control flow changes or value-dependent execution behaviors.

\item \textbf{GroundTruth.} Target performance labels generated through our profiling environment and CMOS PDK analysis are used as benchmark references to compute MAPE.
\end{itemize}

\begin{table*}[htbp]
\caption{Mean Absolute Percentage Error (MAPE) Comparison on Polybench and Modern Workloads with Ablation Study of Progressive Encoding and Dynamic Calibration Strategies.}
\label{tab:polybenchoverall}
\setlength{\tabcolsep}{1pt}
\resizebox{\linewidth}{!}{%
\begin{tabular}{|c|ccccc|ccccc|ccccc|ccccc|}
\hline
\multirow{2}{*}{Benchmark} & \multicolumn{5}{c|}{Static-Power} & \multicolumn{5}{c|}{Static-Area} & \multicolumn{5}{c|}{Static-FF} & \multicolumn{5}{c|}{Dynamic-Cycles} \\ \cline{2-21} 
 & \multicolumn{1}{c|}{NoEnc} & \multicolumn{1}{c|}{Ours} & \multicolumn{1}{c|}{GNNHLS} & \multicolumn{1}{c|}{Tenset} & TLP & \multicolumn{1}{c|}{NoEnc} & \multicolumn{1}{c|}{Ours} & \multicolumn{1}{c|}{GNNHLS} & \multicolumn{1}{c|}{Tenset} & TLP & \multicolumn{1}{c|}{NoEnc} & \multicolumn{1}{c|}{Ours} & \multicolumn{1}{c|}{GNNHLS} & \multicolumn{1}{c|}{Tenset} & TLP & \multicolumn{1}{c|}{NoDPO} & \multicolumn{1}{c|}{Ours} & \multicolumn{1}{c|}{GNNHLS} & \multicolumn{1}{c|}{Tenset} & TLP \\ \hline
adi & 28.7\% & \textcolor{red}{19.4\%} & 64.6\% & 59.3\% & 29.4\% & \textcolor{red}{17.3\%} & 17.7\% & 80.4\% & 65.3\% & 47.2\% & 8.4\% & \textcolor{red}{6.9\%} & 55.1\% & 32.8\% & 27.8\% & 14.4\% & \textcolor{red}{0.6\%} & 42.0\% & 36.5\% & 31.5\% \\ \hline
atax & 22.4\% & \textcolor{red}{5.6\%} & 42.1\% & 34.4\% & 8.5\% & 12.3\% & \textcolor{red}{11.0\%} & 34.4\% & 19.3\% & 25.1\% & 14.8\% & \textcolor{red}{0.2\%} & 28.5\% & 2.0\% & 3.2\% & \textcolor{red}{1.4\%} & 17.5\% & 24.1\% & 20.8\% & 15.6\% \\ \hline
bicg & 28.7\% & \textcolor{red}{20.7\%} & 47.0\% & 60.7\% & 30.8\% & 16.0\% & \textcolor{red} {15.1\%} & 40.6\% & 25.4\% & 20.1\% & 18.4\% & \textcolor{red}{12.2\%} & 30.0\% & 18.0\% & 12.7\% &\textcolor{red}{12.7\%} & 21.1\% & 69.4\% & 45.2\% & 40.0\% \\ \hline
correlation & \textcolor{red}{1.0\%} & 1.6\% & 37.4\% & 38.5\% & 11.1\% & 15.6\% & \textcolor{red}{10.6\%} & 34.4\% & 19.3\% & 11.6\% & 1.9\% & \textcolor{red}{1.0\%} & 9.3\% & 3.2\% & 2.0\% & 12.8\% & \textcolor{red}{0.8\%} & 37.3\% & 23.5\% & 18.3\% \\ \hline
covariance & 28.7\% & \textcolor{red}{5.2\%} & 69.6\% & 45.2\% & 16.0\% & 15.8\% & \textcolor{red}{9.6\%} & 34.4\% & 19.3\% & 10.4\% & \textcolor{red}{2.1\%} & 6.5\% & 10.4\% & 8.7\% & 3.5\% & 13.1\% & \textcolor{red}{0.5\%} & 53.9\% & 30.9\% & 25.6\% \\ \hline
deriche & 28.7\% & \textcolor{red}{20.7\%} & 59.0\% & 60.7\% & 30.7\% & 26.3\% & 20.6\% & 28.1\% & 12.9\% & \textcolor{red}{12.0\%} & 16.0\% & 15.7\% & 27.4\% & 17.9\% & \textcolor{red}{12.7\%} & 23.4\% & \textcolor{red}{9.3\%} & 66.2\% & 45.2\% & 40.0\% \\ \hline
fdtd-2d & 23.7\% & \textcolor{red}{3.7\%} & 53.8\% & 36.3\% & 9.6\% & 19.5\% & \textcolor{red}{7.0\%} & 57.7\% & 42.6\% & 24.2\% & 4.5\% & \textcolor{red}{0.8\%} & 16.6\% & 3.0\% & 2.2\% & 16.6\% & \textcolor{red}{1.5\%} & 30.5\% & 22.4\% & 17.1\% \\ \hline
heat-3d & 22.4\% & \textcolor{red}{5.2\%} & 38.4\% & 34.8\% & 8.8\% & 40.0\% & 25.6\% & 28.1\% & 12.9\% & \textcolor{red}{11.0\%} & 17.2\% & \textcolor{red}{0.1\%} & 21.4\% & 2.2\% & 3.0\% & 36.3\% & \textcolor{red}{13.6\%} & 29.3\% & 21.1\% & 15.9\% \\ \hline
jacobi-2d & 16.2\% & 16.6\% & 11.2\% & 8.8\% & \textcolor{red}{0.1\%} & 41.6\% & 21.0\% & 28.1\% & 12.9\% & \textcolor{red}{11.7\%} & 26.4\% & 12.0\% & \textcolor{red}{8.8\%} & 9.8\% & 15.0\% & 38.9\% & 23.3\% &5.3\% & 72.0\% &  \textcolor{red}{4.8\%} \\ \hline
seidel-2d & 22.4\% & 9.9\% & 21.1\% & 19.8\% & \textcolor{red}{2.2\%} & 41.7\% & 28.3\% & 28.1\% & 12.9\% & \textcolor{red}{12.8\%} & 26.3\% & 23.5\% & 19.8\% & \textcolor{red}{1.8\%} & 3.2\% & 39.2\% & 23.7\% & 25.7\% & 3.8\% & \textcolor{red}{1.2\%} \\ \hline
average(10) & 22.3\% & \textcolor{red}{10.9\%} & 44.4\% & 39.9\% & 14.7\% & 24.6\% & \textcolor{red}{16.7\%} & 39.4\% & 24.3\% & 18.6\% & 13.6\% & \textcolor{red}{7.9\%} & 22.7\% & 10.0\% & 8.5\% & 20.9\% & \textcolor{red}{11.2\%} & 38.4\% & 32.1\% & 21.0\% \\ \hline \hline
Tab. 2-1 & 17.6\% & \textcolor{red}{6.4\%} & 11.8\% & 26.6\% & 20.6\% & 11.2\% &  \textcolor{red}{9.7\%} & 23.6\% & 20.0\% & 24.2\% & 25.1\% & \textcolor{red}{7.7\%} & 27.7\% & 14.0\% & 27.2\% & 16.0\% & \textcolor{red}{6.7\%} & 32.4\% & 9.5\% & 21.8\% \\ \hline
Tab. 2-2 & 13.8\% & \textcolor{red}{11.0\%} & 11.0\% & 25.8\% & 19.6\% & 19.4\% & \textcolor{red}{1.8\%} & 21.5\% & 17.9\% & 23.6\% & 25.0\% & \textcolor{red}{2.9\%} & 22.4\% & 8.6\% & 21.6\% & 15.8\% & \textcolor{red}{7.0\%} & 32.5\% & 9.3\% & 18.0\% \\ \hline
Tab. 2-3 & 19.4\% & 13.3\% & \textcolor{red}{9.2\%} & 24.0\% & 18.8\% & 16.8\% & \textcolor{red}{2.4\%} & 16.7\% & 13.1\% & 18.9\% & 13.6\% & \textcolor{red}{1.0\%} & 15.2\% & 1.4\% & 14.8\% & 20.4\% & \textcolor{red}{3.6\%} & 20.3\% & 21.6\% & 5.5\% \\ \hline
Tab. 2-4 & 24.5\% & 8.5\% & 6.1\% & 8.8\% & \textcolor{red}{ 4.9\%} & \textcolor{red}{0.7\%} & 17.0\% & 67.6\% & 71.2\% & 65.3\% &\textcolor{red}{ 5.4\%} & 46.2\% & 50.2\% & 64.0\% & 51.4\% & 36.8\% & 13.8\% & \textcolor{red}{1.5\%} & 40.3\% & 17.6\% \\ \hline
Tab. 2-5 & 16.7\% & 11.0\% & \textcolor{red}{6.7\%} & 21.5\% & 16.0\% & 16.1\% & \textcolor{red}{6.0\%} & 10.5\% & 16.9\% & 13.0\% & 18.4\% & 4.3\% & 9.9\% &\textcolor{red}{3.9\%} & 9.6\% & 12.8\% & \textcolor{red}{10.0\%} & 25.8\% & 16.1\% & 11.1\% \\ \hline
Tab. 2-6 & 46.8\% & \textcolor{red}{11.5\%} & 56.1\% & 41.3\% & 45.7\% & \textcolor{red}{10.0\%} & 18.3\% & 26.6\% & 30.2\% & 24.4\% & 61.5\% &\textcolor{red}{ 31.3\%} & 57.4\% & 71.2\% & 57.4\% & 91.8\% & 69.0\% & \textcolor{red}{52.8\%} & 94.7\% & 67.3\% \\ \hline
Tab. 2-7 & 23.8\% & 10.8\% & 5.8\% & 9.0\% &\textcolor{red}{ 3.6\%} & \textcolor{red}{7.9\%} & 15.5\% & 17.3\% & 20.9\% & 15.1\% & 26.4\% &\textcolor{red}{ 23.3\% }& 23.4\% & 37.2\% & 23.9\% & 57.5\% & 34.8\% & \textcolor{red}{19.2\%} & 61.1\% & 34.1\% \\ \hline
Tab. 2-8 & 32.8\% & 14.9\% & 18.4\% & \textcolor{red}{3.6\%} & 9.2\% & \textcolor{red}{5.9\%} & 16.9\% & 21.0\% & 24.6\% & 19.2\% &\textcolor{red}{ 24.0\%} & 35.5\% & 37.2\% & 51.0\% & 37.4\% & 55.4\% & 32.6\% & \textcolor{red}{16.8\%} & 58.7\% & 31.6\% \\ \hline
Tab. 2-9 & 17.8\% & \textcolor{red}{3.7\%} & 8.1\% & 22.9\% & 17.6\% & 12.4\% & \textcolor{red}{1.7\%} & 21.7\% & 18.1\% & 24.3\% & 24.9\% & \textcolor{red}{2.3\%} & 19.7\% & 5.9\% & 19.1\% & \textcolor{red}{9.6\%} & 13.4\% & 29.6\% & 12.2\% & 18.1\% \\ \hline
Tab. 2-10 & 12.9\% & 13.4\% & \textcolor{red}{1.7\%} & 13.1\% & 7.5\% & 15.4\% & \textcolor{red}{6.3\%} & 17.3\% & 13.7\% & 19.7\% & 21.9\% & 11.2\% & 23.4\% &\textcolor{red}{ 9.6\%} & 23.0\% &\textcolor{red}{ 10.4\%} & 12.3\% & 29.1\% & 12.8\% & 14.7\% \\ \hline
Tab. 2-11 & 42.1\% & \textcolor{red}{3.3\%} & 85.7\% & 70.9\% & 76.7\% &10.3\% & \textcolor{red}{7.9\%} & 18.5\% & 22.1\% & 16.6\% & 15.7\% & \textcolor{red}{5.8\%} & 7.9\% & 5.9\% & 8.0\% & 16.2\% & \textcolor{red}{6.6\%} & 23.1\% & 18.7\% & 9.2\% \\ \hline
Tab. 2-12 & 26.1\% & 10.9\% & \textcolor{red}{5.3\%} & 20.1\% & 14.4\% & \textcolor{red}{7.7\%} & 13.0\% & 25.2\% & 28.8\% & 22.9\% & 28.2\% & 10.2\% & 17.8\% &\textcolor{red}{ 4.0\%} & 17.1\% & 17.6\% & \textcolor{red}{5.2\%} & 34.0\% & 7.9\% & 20.0\% \\ \hline
Tab. 2-13 & 24.7\% & \textcolor{red}{8.4\%} & 24.1\% & 9.3\% & 10.3\% & 15.8\% & 13.0\% & \textcolor{red}{10.3\%} & 26.7\% & 12.2\% & 16.8\% & \textcolor{red}{0.8\%} & 5.8\% & 7.9\% & 4.7\% & 14.9\% & \textcolor{red}{7.3\%} & 24.3\% & 17.6\% & 13.2\% \\ \hline
Tab. 2-14 & \textcolor{red}{13.6\%} & 15.3\% & 14.0\% & 28.8\% & 24.0\% & 17.5\% & \textcolor{red}{5.5\%} & 31.3\% & 27.7\% & 33.4\% & 39.1\% & \textcolor{red}{16.4\%} & 30.7\% & 26.9\% & 36.6\% & 29.5\% &6.9\% & 46.2\% & \textcolor{red}{4.3\%} & 34.9\% \\ \hline
average(14) & 23.7\% & \textcolor{red}{10.2\%} & 18.9\% & 23.3\% & 20.6\% & 11.9\% &\textcolor{red}{9.6\%} & 23.5\% & 25.1\% & 23.8\% & 24.7\% & \textcolor{red}{14.2\%} & 24.9\% & 22.2\% & 25.1\% & 28.9\% & \textcolor{red}{16.4\%} & 27.7\% & 27.5\% & 22.7\% \\ \hline \hline
TPU        & 11.5\% & \textcolor{red}{9.8\%}  & 12.9\% & 12.8\% & 10.4\% & 12.3\% & \textcolor{red}{10.2\%} & 10.4\% & 17.0\% & 17.8\% & 11.9\% & \textcolor{red}{8.2\%} & 14.5\% & 11.4\% & 20.1\% & 13.4\% & \textcolor{red}{10.3\%} & 13.8\% & 12.2\% & 14.0\% \\ \hline
Eyeriss    & 11.0\% & 11.0\% & 10.9\% & 13.0\% & \textcolor{red}{10.7\%} & 12.9\% & \textcolor{red}{9.1\%}  & 14.3\% & 12.8\% & 15.8\% & 8.4\%  & \textcolor{red}{8.3\%}  & 14.7\% & 8.4\%  & 20.3\% & 10.7\% & \textcolor{red}{9.6\%}  & 11.8\% & 13.0\% & 12.0\% \\ \hline
Shidiannao & 10.7\% & \textcolor{red}{9.2\%}  & 10.4\% & 12.5\% & 10.0\% & 12.3\% & \textcolor{red}{10.1\%} & 12.4\% & 10.2\% & 15.3\% & \textcolor{red}{6.9\%}  & 9.1\%  & 10.4\% & 10.5\% & 20.6\% & 12.0\% & \textcolor{red}{10.0\%} & 14.3\% & 15.5\% & 11.4\% \\ \hline
\end{tabular}%
}
\end{table*}

\subsection{Comprehensive Evaluation}\label{sec:comprehensiveval}

\subsubsection*{Accuracy Evaluation} We evaluated LLMulator across 24 workloads (10 Polybench kernels and 14 modern applications) covering image processing and NLP domains. As shown in Table ~\ref{tab:polybenchoverall}, LLMulator achieves an average MAPE of 12.2\% across all metrics (static power/area/FF and dynamic cycles), outperforming GNNHLS (28.9\% MAPE) and TLP (20.0\% MAPE) by 16.7\% and 7.8\% respectively, which shows better application generalization. Notably for dynamic cycle prediction, our dynamic calibration framework reduces MAPE from initial 28.9\% to 16.4\% after 5 DPO iterations, demonstrating 41\% accuracy improvement over static models. This highlights the enhanced input-aware generalization and the effectiveness of the methods discussed in Section~\ref{sec:dynamicmethod}. We validate LLMulator's effectiveness on dynamic input-depended control flow against methods like Tenset-MLP~\cite{zheng2021tenset}, which extract profiling-derived dynamic features to realize limited input adaptivity. As shown in Table ~\ref{tab:polybenchoverall}, LLMulator achieves a superior MAPE  versus Tenset-MLP because Tenset-MLP only make prediction based on simple parameters like loop range but not real inputs.

The numeric modeling-based static prediction shows particular strength in edge-case scenarios. For extreme values (>95\% in training data), traditional regression methods like TLP exhibit higher error compared to our digit-wise classification through confidence-based beam search. This is evidenced in T5-based workload (Table ~\ref{tab:benchmarkanalysis}-12) where LLMulator achieves  10.9\% MAPE for power prediction versus TLP's 14.4\%. 

\subsubsection*{Dealing with Errors} Some workloads, such as jacobi-2d, exhibit higher prediction errors due to their elevated structural and semantic complexity. These programs often include deeply nested class hierarchies, repeated invocations of user-defined data structures (\emph{e.g.,} the \texttt{FeatureMap} class), and complex control/data interactions. Such factors increase the difficulty of program analysis and modeling by LLMs struggling with deeply abstracted or non-local program semantics. This limitation is not specific to LLMulator but reflects a current open challenge in applying LLMs to nuanced program understanding~\cite{hu2025dynacodedynamiccomplexityawarecode,chatterjee2024phaedrus}. Future work will explore program normalization techniques and LLM preconditioning strategies to make LLMulator more accurate.

\begin{table}[htbp]
\setlength{\tabcolsep}{0.5pt}
\caption{LLMulator Runtime Latency (Seconds) of Prediction Models on Polybench Benchmarks: LLMulator vs. Baseline Approaches.}
\label{tab:predictioncomp}
\resizebox{\linewidth}{!}{%
\begin{tabular}{|c|c|c|c|c|c|c|c|c|c|c|}
\hline
 & heat-3d & corre. & jacobi-2d & fdtd-2d & deriche & seidel-2d & adi & covar. & bicg & atax \\ \hline
GNNHLS & 0.12 & 0.12 & 0.11 & 0.11 & 0.11 & 0.11 & 0.11 & 0.11 & 0.11 & 0.11 \\ \hline
Tenset & 0.08 & 0.08 & 0.08 & 0.08 & 0.08 & 0.08 & 0.08 & 0.08 & 0.08 & 0.08 \\ \hline
TLP & 0.21 & 0.08 & 0.10 & 0.15 & 0.10 & 0.09 & 0.13 & 0.10 & 0.09 & 0.07 \\ \hline
Ours & 1.45 & 0.86 & 0.84 & 1.09 & 0.95 & 0.72 & 1.36 & 0.97 & 0.95 & 1.06 \\ \hline
\end{tabular}%
}
\end{table}

\begin{table}[htbp]
\caption{LLMulator Runtime Latency Comparison (Seconds) for Cycle Predictions: LLMulator without Dynamic Prediction Acceleration vs. LLMulator for Table ~\ref{tab:benchmarkanalysis} Workloads.}
\setlength{\tabcolsep}{1pt}
\label{tab:acceleratingcmompare}
\resizebox{\linewidth}{!}{%
\begin{tabular}{|l|l|l|l|l|l|l|l|l|l|l|l|l|l|l|}
\hline
Tab. ~\ref{tab:benchmarkanalysis}-Index & 1 & 2 & 3 & 4 & 5 & 6 & 7 & 8 & 9 & 10 & 11 & 12 & 13 & 14 \\ \hline
NoAccel& 1.01 & 1.62 & 1.02 & \textcolor{red}{0.95} & 1.01 & 1.62 & 0.99 & 1.10 & 1.38 & 1.43 & 1.53 & 1.54 & 1.01 & 0.99 \\ \hline
HasAccel& \textcolor{red}{0.93} & \textcolor{red}{1.04} & \textcolor{red}{0.94} & 1.01 & \textcolor{red}{0.95} & \textcolor{red}{1.12} & \textcolor{red}{0.96} & \textcolor{red}{1.03} & \textcolor{red}{1.05} & \textcolor{red}{1.08} & \textcolor{red}{1.07} & \textcolor{red}{0.97} & \textcolor{red}{0.99} & \textcolor{red}{0.96} \\ \hline
\end{tabular}%
}
\end{table}
\subsubsection*{Runtime Evaluation} As shown in Table~\ref{tab:predictioncomp}, LLMulator’s average runtime latency on Polybench (1.01s per prediction) is higher than GNNHLS (0.11s) due to LLM computation overhead. However, our dynamic acceleration techniques narrow this gap, reducing the average runtime from 1.23s to 1.00s  in Table~\ref{tab:acceleratingcmompare}, achieved through selective attention masking (Section~\ref{sec:dynamicaccel}). This runtime overhead is acceptable compared to the longer synthesis times associated with EDA tools.

\subsubsection*{Comparing with Rule-based methods}
\begin{figure}[!htbp] 
    \centering 
    
    \begin{minipage}[t]{0.75\linewidth} 
        \centering 
            \centering
            \includegraphics[width=\linewidth]{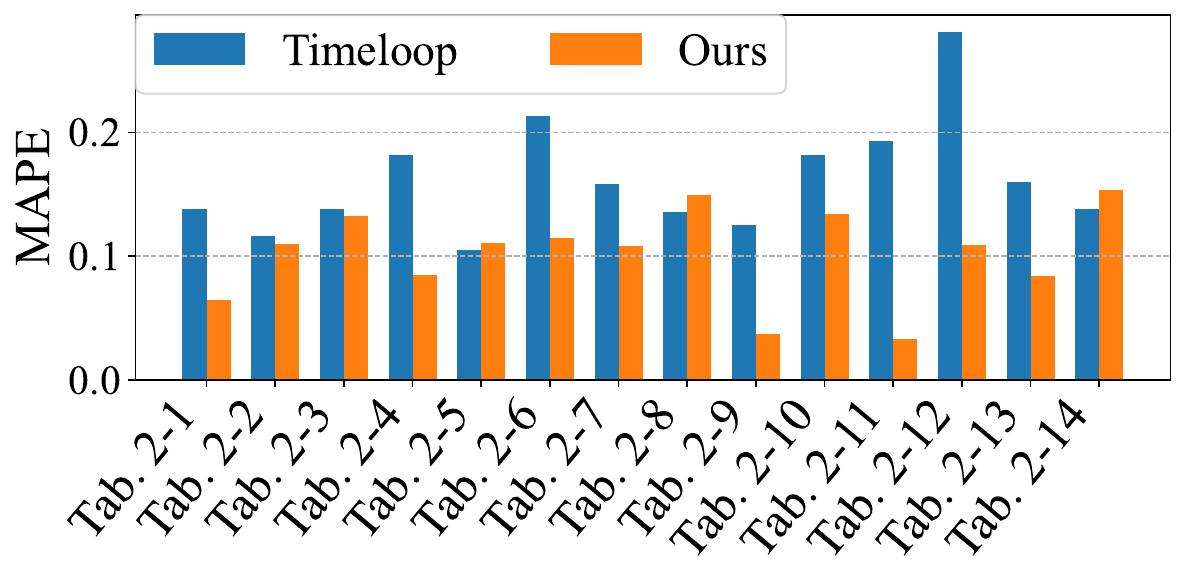}

            \caption{Compare with Timeloop MAPE.} 
            \label{fig:timeloopcomp} 
    \end{minipage}
    \hfill 

\end{figure}

We compare LLMulator to Timeloop, a rule-based tensor algebra-only simulator, with our framework using deep learning operators (\emph{i.e.,} operators supported by Timeloop) using workloads in  Table \ref{tab:benchmarkanalysis} to estimate power. As Figure~\ref{fig:timeloopcomp} shows, LLMulator achieves a lower MAPE of 10.2\% versus Timeloop's 16.2\% on average. This improvement reflects LLMulator's superior data-fitting capability and, more importantly, its generalizability to complex HLS applications beyond the expressiveness of static rule-based approaches. Timeloop is fundamentally limited to evaluating regular, loop-nest-based tensor computations. It cannot natively model workloads with control flow variability, I/O interactions, or heterogeneous operator sequences as LLMulator. Handling such cases requires manually decomposing the program into atomic tensor operators and externally aggregating their metrics, leading to reduced modeling fidelity.

\subsection{Ablation \& Sensitivity Study}\label{sec:expeablation}

\subsubsection*{Effectiveness of Numeric Modeling-based Static Prediction} To evaluate the effectiveness of the progressive program encoding, we present the results in Table~\ref{tab:polybenchoverall}. Without the proposed encoding strategy, the static metrics average MAPEs are 23.7\%, 11.9\%, and 24.7\% for static predictions. With the encoding approach, the average MAPEs reduce to 10.2\%, 9.6\%, and 14.2\%, as shown in Table~\ref{tab:polybenchoverall}. This consistent increase demonstrates the effectiveness of the proposed method. Additionally, we assess the output numeric modeling method, which relates to the error in MSE with logits as confidence in Table~\ref{tab:confrelation}. It reveals a Pearson correlation coefficient of -0.44 between prediction confidence (final-layer logits) and MSE, establishing quantifiable uncertainty estimates. This indicates that the proposed output numerical modeling reflects the confidence of the predictions.

\begin{table}[htbp]
\caption{Correlation Analysis of Mean Squared Error (MSE) and Prediction Confidence (Logit) for Flip-Flop (FF) Estimates on Randomly Sampled Workloads.}
\label{tab:confrelation}
\setlength{\tabcolsep}{3pt}
\resizebox{\linewidth}{!}{%
\begin{tabular}{|l|l|l|l|l|l|l|l|l|l|l|l|l|}
\hline
\textbf{Confi} & 0.47 & 0.43 & 0.74 & 0.22 & 0.20 & 0.74 & 0.77 & 0.30 & 0.22 & 0.61 & 0.58 & 0.33 \\ \hline
\textbf{Pred} & 7 & 18 & 18 & 17 & 9 & 3 & 1 & 8 & 24 & 0 & 0 & 11 \\ \hline
\textbf{Real} & 5 & 4 & 14 & 44 & 8 & 6 & 1 & 22 & 18 & 1 & 2 & 12 \\ \hline
\textbf{MSE} & 4 & 196 & 16 & 729 & 1 & 9 & 0 & 196 & 36 & 1 & 4 & 1 \\ \hline
\end{tabular}%
}
\end{table}

\begin{table}[htbp]
\caption{Ablation Study of Progressive Data Synthesis Components on Prediction Accuracy (MAPE) Across Hardware Metrics.}
\label{tab:datasyncomparision}
\resizebox{\linewidth}{!}{%
\begin{tabular}{|c|cc|cc|cc|cc|}
\hline
 & \multicolumn{2}{c|}{\textbf{Power}} & \multicolumn{2}{c|}{\textbf{Area}} & \multicolumn{2}{c|}{\textbf{FF}} & \multicolumn{2}{c|}{\textbf{Cycles}} \\ \hline
 & \multicolumn{1}{c|}{\textbf{No-A}} & \textbf{All} & \multicolumn{1}{c|}{\textbf{No-A}} & \textbf{All} & \multicolumn{1}{c|}{\textbf{No-A}} & \textbf{All} & \multicolumn{1}{c|}{\textbf{No-A}} & \textbf{All} \\ \hline
Tab. 2-1 & \multicolumn{1}{c|}{9.3\%} & \textcolor{red}{6.4\%} & \multicolumn{1}{c|}{44.7\%} & \textcolor{red}{9.7\%} & \multicolumn{1}{c|}{11.9\%} & \textcolor{red}{7.7\%} & \multicolumn{1}{c|}{37.6\%} & \textcolor{red}{21.8\%} \\ \hline
Tab. 2-2 & \multicolumn{1}{c|}{23.6\%} & \textcolor{red}{11.0\%} & \multicolumn{1}{c|}{43.1\%} & \textcolor{red}{1.8\%} & \multicolumn{1}{c|}{5.2\%} & \textcolor{red}{2.9\%} & \multicolumn{1}{c|}{35.6\%} & \textcolor{red}{18.0\%} \\ \hline
Tab. 2-3 & \multicolumn{1}{c|}{16.5\%} & \textcolor{red}{13.3\%} & \multicolumn{1}{c|}{36.9\%} & \textcolor{red}{2.4\%} & \multicolumn{1}{c|}{7.7\%} & \textcolor{red}{1.0\%} & \multicolumn{1}{c|}{20.2\%} & \textcolor{red}{5.5\%} \\ \hline
Tab. 2-4 & \multicolumn{1}{c|}{18.6\%} & \textcolor{red}{8.5\%} & \multicolumn{1}{c|}{47.0\%} & \textcolor{red}{17.0\%} & \multicolumn{1}{c|}{67.8\%} & \textcolor{red}{46.2\%} & \multicolumn{1}{c|}{22.5\%} & \textcolor{red}{17.6\%} \\ \hline
Tab. 2-5 & \multicolumn{1}{c|}{\textcolor{red}{6.6\%}} & 11.0\% & \multicolumn{1}{c|}{30.6\%} & \textcolor{red}{6.0\%} & \multicolumn{1}{c|}{8.6\%} & \textcolor{red}{4.3\%} & \multicolumn{1}{c|}{23.3\%} & \textcolor{red}{11.1\%} \\ \hline
Tab. 2-6 & \multicolumn{1}{c|}{70.3\%} & \textcolor{red}{11.5\%} & \multicolumn{1}{c|}{\textcolor{red}{6.7\%}} & 18.3\% & \multicolumn{1}{c|}{81.7\%} & \textcolor{red}{31.3\%} & \multicolumn{1}{c|}{\textcolor{red}{51.4\%}} & 67.3\% \\ \hline
Tab. 2-7 & \multicolumn{1}{c|}{20.1\%} & \textcolor{red}{10.8\%} & \multicolumn{1}{c|}{\textcolor{red}{2.1\%}} & 15.5\% & \multicolumn{1}{c|}{40.5\%} & \textcolor{red}{23.3\%} & \multicolumn{1}{c|}{37.7\%} & \textcolor{red}{34.1\%} \\ \hline
Tab. 2-8 & \multicolumn{1}{c|}{32.2\%} & \textcolor{red}{14.9\%} & \multicolumn{1}{c|}{\textcolor{red}{0.4\%}} & 16.9\% & \multicolumn{1}{c|}{59.3\%} & \textcolor{red}{35.5\%} & \multicolumn{1}{c|}{\textcolor{red}{12.9\%}} & 31.6\% \\ \hline
Tab. 2-9 & \multicolumn{1}{c|}{\textcolor{red}{3.0\%}} & 3.7\% & \multicolumn{1}{c|}{42.9\%} & \textcolor{red}{1.7\%} & \multicolumn{1}{c|}{2.4\%} & \textcolor{red}{2.3\%} & \multicolumn{1}{c|}{34.4\%} & \textcolor{red}{18.1\%} \\ \hline
Tab. 2-10 & \multicolumn{1}{c|}{15.8\%} & \textcolor{red}{13.4\%} & \multicolumn{1}{c|}{38.0\%} & \textcolor{red}{6.3\%} & \multicolumn{1}{c|}{\textcolor{red}{2.0\%}} & 11.2\% & \multicolumn{1}{c|}{35.0\%} & \textcolor{red}{14.7\%} \\ \hline
Tab. 2-11 & \multicolumn{1}{c|}{97.9\%} & \textcolor{red}{3.3\%} & \multicolumn{1}{c|}{\textcolor{red}{0.7\%}} & 7.9\% & \multicolumn{1}{c|}{13.9\%} & \textcolor{red}{5.8\%} & \multicolumn{1}{c|}{27.1\%} & \textcolor{red}{9.2\%} \\ \hline
Tab. 2-12 & \multicolumn{1}{c|}{\textcolor{red}{7.5\%}} & 10.9\% & \multicolumn{1}{c|}{\textcolor{red}{3.7\%}} & 13.0\% & \multicolumn{1}{c|}{\textcolor{red}{4.8\%}} & 10.2\% & \multicolumn{1}{c|}{37.8\%} & \textcolor{red}{20.0\%} \\ \hline
Tab. 2-13 & \multicolumn{1}{c|}{35.7\%} & \textcolor{red}{8.4\%} & \multicolumn{1}{c|}{31.2\%} & \textcolor{red}{13.0\%} & \multicolumn{1}{c|}{11.7\%} & \textcolor{red}{0.8\%} & \multicolumn{1}{c|}{24.3\%} & \textcolor{red}{13.2\%} \\ \hline
Tab. 2-14 & \multicolumn{1}{c|}{\textcolor{red}{0.1\%}} & 15.3\% & \multicolumn{1}{c|}{52.2\%} & \textcolor{red}{5.5\%} & \multicolumn{1}{c|}{18.4\%} & \textcolor{red}{16.4\%} & \multicolumn{1}{c|}{44.4\%} & \textcolor{red}{34.9\%} \\ \hline
average & \multicolumn{1}{c|}{25.5\%} & \textcolor{red}{10.2\%} & \multicolumn{1}{c|}{27.1\%} & \textcolor{red}{9.6\%} & \multicolumn{1}{c|}{24.0\%} & \textcolor{red}{14.2\%} & \multicolumn{1}{c|}{31.7\%} & \textcolor{red}{22.7\%} \\ \hline
\end{tabular}%
}
\end{table}

\subsubsection*{Effectiveness of Dataset Synthesizer}
Since the previous experiment focused on using the same dataset, we conduct an ablation study on the dataset, as shown in Table~\ref{tab:datasyncomparision}, to systematically validate the effectiveness of the proposed data synthesizer. When data augmentation is disabled, except for AST-based data and direct data format (No-A columns), the model shows significant performance degradation across all cost prediction tasks. In contrast, the complete framework (All columns) achieves an average MAPE reduction from 27.1\% to 14.2\%, demonstrating the effectiveness of the proposed dataset augmentation method. 
Specifically, for unconventional dataflow patterns, such as the GAN-based super-resolution workload (in Table~\ref{tab:benchmarkanalysis}-6), data augmentation reduces the FF prediction MAPE by 50.4\% (from 81.7\% to 31.3\%). Through LLM-guided semantic-preserving mutations (\emph{e.g.,} replacing 3$\times$3 convolutions with 5$\times$5 depthwise variants), we effectively expand the model's coverage of irregular computation graphs.  In addition, we apply the data synthesizer enhancement to the other baseline methods to evaluate its effectiveness.  Table \ref{tab:datasynmapediff_1} shows the MAPE using the new synthesized dataset versus the original dataset, where negative values indicate that the MAPE without the synthesized data is higher than with it. Compared to their original datasets, GNNHLS, TLP, and Tenset-MLP exhibit average MAPE reductions from 33.2\%, 29.9\%, and 34.0\% down to 27.5\%, 22.7\%, and 27.7\%, respectively, after adding the synthesized data.

\begin{figure}[htbp]
    \centering
    \includegraphics[width=\linewidth]{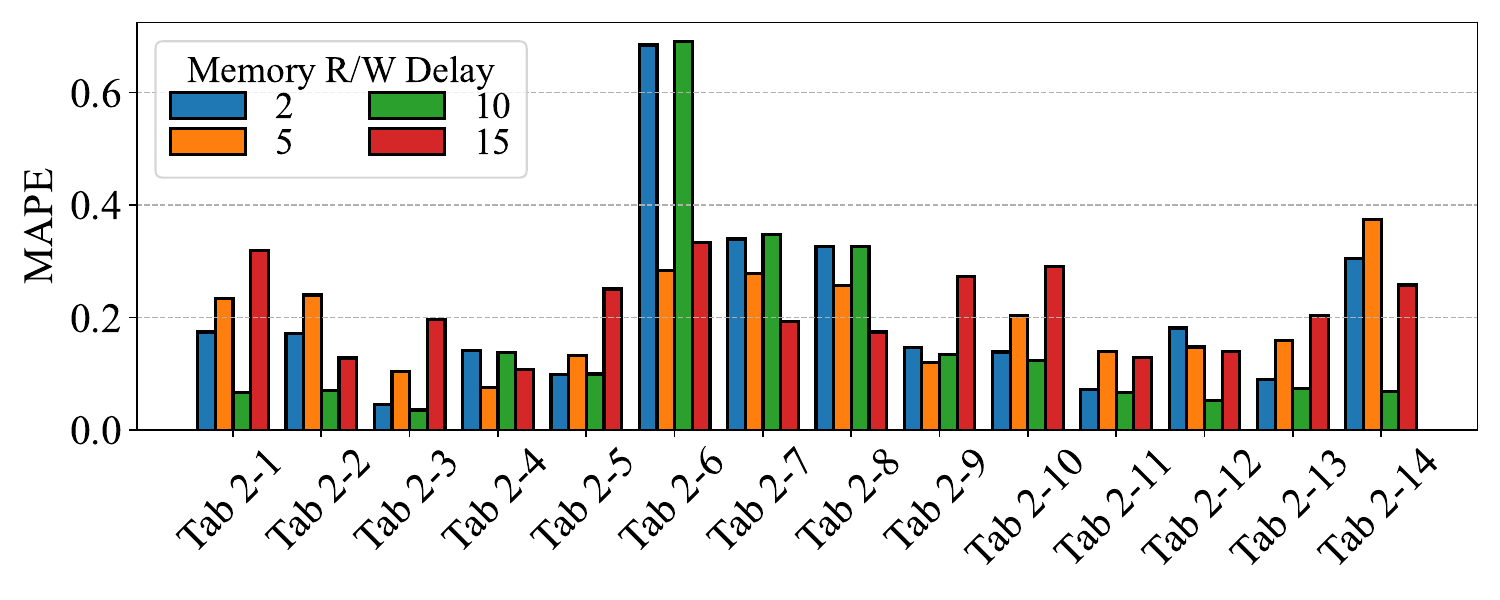}
    \caption{Cycle Prediction Error Distribution Across Memory Latency Parameters.}
    \label{fig:memorycyclemape}
\end{figure}

To systematically evaluate the effectiveness of hardware generalization, we vary the memory read and write delay across values of 2, 5, 10, and 15 to assess MAPE across the benchmark. The evaluation is conducted in terms of cycles, as shown in Figure ~\ref{fig:memorycyclemape}. Specifically, for cycles 2, 5, 10, and 15, the average MAPE values are 20.8\%, 19.6\%, 16.4\%, and 21.4\%, respectively. Notably, there is no significant increase in error for cycle 15 compared to the other cycles, even though it falls outside the data synthesizer’s parameters. Based on these results, we conclude that the dataset synthesizer and dynamic calibration enable LLMulator to adapt effectively to different hardware parameters. 

\begin{table}[ht]
\centering
\caption{MAPE Difference with and without the proposed Data Synthesizer.}
\label{tab:datasynmapediff_1}
\resizebox{\linewidth}{!}{%
\begin{tabular}{|l|l|l|l|l|l|l|l|}
\hline
       & 1       & 2       & 3       & 4       & 5       & 6      & 7       \\ \hline
Tenset & 6.4\%   & -26.5\% & 7.0\%   & -13.0\% & -13.0\% & 7.0\%  & -13.0\% \\ \hline
TLP    & -14.4\% & -18.5\% & -18.6\% & 12.2\%  & -18.6\% & 18.4\% & 18.8\%  \\ \hline
GNNHLS & -11.0\% & -8.7\%  & -16.2\% & -46.2\% & -4.5\%  & -5.3\% & -1.6\%  \\ \hline \hline
       & 8      & 9       & 10      & 11      & 12      & 13      & 14       \\ \hline
Tenset & 7.0\%  & 1.2\%   & -5.8\%  & -3.0\%  & -13.8\% & -2.0\%  & -18.7\%  \\ \hline
TLP    & 18.6\% & -15.4\% & -18.2\% & -17.8\% & -17.9\% & -15.0\% & -15.1\%  \\ \hline
GNNHLS & -0.8\% & 18.4\%  & -12.4\% & 11.8\%  & 8.6\%   & -5.8\%  & -14.5\%  \\ \hline
\end{tabular}
}
\end{table}

\begin{table}[htbp]
\caption{Impact of Data Dependency Length on LLMulator Runtime Latency (Seconds) with Dynamic Prediction Acceleration.}
\setlength{\tabcolsep}{1pt}
\label{tab:acceleratecomp}
\resizebox{\linewidth}{!}{%
\begin{tabular}{|l|llllllllllllll|}
\hline
DataDepLen & 0 & 564 & 554 & 544 & 534 & 524 & 514 & 504 & 494 & 484 & 474 & 464 & 454 & 444 \\ \hline
DataLength & 1524 & 2088 & 2078 & 2068 & 2058 & 2048 & 2038 & 2028 & 2018 & 2008 & 1998 & 1988 & 1978 & 1968 \\ \hline
NoOptTime & 1.51 & 1.27 & 1.39 & 1.62 & 1.11 & 1.17 & 1.44 & 1.22 & 1.44 & 1.29 & 1.09 & 1.11 & 1.11 & 1.11 \\ \hline
OptSpeed & 1.13 & 1.18 & 1.19 & 1.10 & 1.03 & 1.05 & 1.00 & 1.02 & 0.98 & 1.01 & 1.00 & 0.99 & 0.99 & 1.12 \\ \hline \hline
DataDepLen & 434 & 424 & 414 & 404 & 394 & 384 & 374 & 364 & 354 & 344 & 334 & 324 & 314 & 304 \\ \hline
DataLength & 1958 & 1948 & 1938 & 1928 & 1918 & 1908 & 1898 & 1888 & 1878 & 1868 & 1858 & 1848 & 1838 & 1828 \\ \hline
NoOptTime & 1.09 & 4.18 & 1.13 & 1.21 & 1.23 & 1.04 & 1.25 & 1.21 & 1.06 & 1.20 & 1.35 & 0.99 & 1.27 & 1.03 \\ \hline
OptTime & 0.97 & 1.55 & 1.01 & 1.04 & 0.97 & 1.13 & 1.09 & 0.95 & 1.03 & 1.06 & 1.25 & 0.94 & 0.96 & 0.92 \\ \hline
\end{tabular}%
}
\end{table}

\subsubsection*{Ablation study on Predicting Speed} As shown in Table ~\ref{tab:acceleratingcmompare}, selective attention masking reduces prediction latency by from 1.23s to 1.01s on average. The acceleration effect remains consistent across varying data dependency lengths (Table \ref{tab:acceleratecomp}), demonstrating our approach's adaptability to different control flow complexities. 
The systematic evaluation in Table \ref{tab:acceleratecomp} demonstrates that our acceleration strategy maintains stable performance across varying data dependency lengths, where \texttt{DataDepLen} refers to the input-dependent operator byte length, and \texttt{DataLength} represents the total dataflow text byte length. When the dataflow length expands from 1524 to 1828, the optimized scheme achieves a standard deviation of merely 0.13s. Notably, in extreme-length scenarios (\texttt{DataDepLen}=514), attention sparsification reduces runtime by 30.6\%, conclusively demonstrating that the partial attention mechanism proposed in Section \ref{sec:dynamicaccel} effectively addresses context length challenges.

\subsubsection*{Ablation Study Summary} Our ablation study (Tables ~\ref{tab:polybenchoverall} and \ref{tab:datasyncomparision}) reveals the individual impact of LLMulator’s core modules. Numeric encoding lowers the average MAPE of static power, area, and FF by 8.8 percentage points; Dynamic calibration via DPO reduces dynamic cycle‐count error by 9.7 percentage points; and the dataset synthesizer delivers the greatest gain, achieving a 12.9 percentage‐point average error reduction across all tasks. 

\subsubsection*{Sensitivity on Base Model Size}
We evaluate the impact of varying model parameters on prediction accuracy by training several base models, including LLaMA-3.2 1B, LLaMA-3.1 8B, and Qwen2.5-0.5B ~\cite{qwen2}, all with identical training parameters and data. As shown in Table ~\ref{tab:modelsizecomp}, we compare with the workloads predicted cycles with workloads in  Table ~\ref{tab:benchmarkanalysis}, where MAPE are 22.9\%, 16.4\%, 15.3\% on average in 0.5B, 1B and 8B model, suggesting that larger models yield more reliable performance estimates under identical training settings.

\begin{table}[]
\caption{Comparison of MAPE for Cycles Prediction at Different Model Scales.}
\setlength{\tabcolsep}{0.5pt}
\label{tab:modelsizecomp}
\resizebox{\linewidth}{!}{%
\begin{tabular}{|l|l|l|l|l|l|l|l|l|l|l|l|l|l|l|}
\hline
Tab. 2 & 1      & 2      & 3      & 4      & 5      & 6      & 7      & 8      & 9      & 10     & 11    & 12     & 13     & 14     \\ \hline
0.5B    & 12.1\% & 10.8\% & 11.5\% & 29.9\% & 5.6\%  & \redtext{51.2\%} & 65.3\% & 61.4\% & 15.4\% & 16.2\% & 8.3\% & 10.5\% & 20.3\% & \redtext{2.2\%}  \\ \hline
1B     & \redtext{6.7\%}  & \redtext{7.0\%}  & 3.6\%  & 13.8\% & 10.0\% & 69.0\% & \redtext{34.8\%} & \redtext{32.6\%} & 13.4\% & 12.3\% & 6.6\% & \redtext{5.2\%}  & 7.3\%  & 6.9\%  \\ \hline
8B     & 8.3\%  & 8.2\%  & \redtext{2.6\%}  & \redtext{3.7\%}  & \redtext{1.8\%}  & 76.7\% & 43.2\% & 40.9\% & \redtext{6.0\%}  & \redtext{5.1\%}  & \redtext{1.4\%} & 10.7\% & \redtext{0.5\%}  & 5.2\%  \\ \hline
\end{tabular}%
}
\end{table}

\subsection{Real World Case Study}\label{sec:realworld}
To demonstrate the applicability of LLMulator to real-world accelerator designs, we evaluate its prediction accuracy on representative ASIC-style dataflow architectures, including Google TPU v1 (weight-stationary) ~\cite{tpu}, Eyeriss (input-stationary) ~\cite{chen2016eyeriss}, and ShiDianNao (output-stationary) ~\cite{du2015shidiannao}.

\paragraph{Real World Dataflow Accelerators}
We begin with TPU v1~\cite{tpu}, a canonical weight-stationary dataflow design. To mimic alternative dataflow strategies, we modify the loop unrolling pattern: converting to input-stationary (Eyeriss-like) ~\cite{chen2016eyeriss} and output-stationary (ShiDianNao-like) ~\cite{du2015shidiannao} variants. These loop schedule variants are synthetically compiled from PolyBench suite (Gemm workload), with their corresponding hardware mappings adjusted accordingly. We evaluate prediction accuracy using a pre-trained LLMulator model without fine-tuning on these specific designs.

As shown in the last three rows of Table~\ref{tab:polybenchoverall}, LLMulator achieves MAPE between 6.9\% and 10.7\%, outperforming prior methods (\emph{e.g.,} GNNHLS, Tenset-MLP) in most configurations. This result affirms LLMulator’s robustness and transferability to new dataflow mappings and hardware styles, even without explicit retraining.


\paragraph{Real World Dataflow Applications}
To further assess LLMulator's effectiveness on realistic workloads, we evaluate it on C-like dataflow programs compiled for TPU execution using MLIRSynth ~\cite{mlirsynth}. We dynamically calibrate LLMulator using input profiles collected during TPU runs and generate ground-truth execution profiles by averaging the lowest-latency results across ten independent executions.

We compare LLMulator against TLP and Tenset-MLP baselines that leverge the profiles for modeling. As shown in Table \ref{tab:dataflowapp}, LLMulator achieves superior accuracy across a wide set of PolyBench applications compiled for TPU. These results validate LLMulator’s input-adaptive prediction ability and show its strong generalization performance on real programmable dataflow accelerators.


\begin{table}[htbp]
\setlength{\tabcolsep}{0.7pt}
\caption{Dataflow Application MAPE on Polybench.}
\label{tab:dataflowapp}
\resizebox{\linewidth}{!}{%
\begin{tabular}{|l|l|l|l|l|l|l|l|l|l|l|}
\hline
       & adi    & atax   & bicg   & corre. & covar. & deriche & fdtd-2d & heat-3d & jacobi. & seidel. \\ \hline
Ours   & \redtext{6.2\%}  & \redtext{15.3\%} & 18.0\% & \redtext{1.6\%}  & \redtext{2.6\%}  & \redtext{12.2\%}  & 10.5\%  & 17.5\%  & \redtext{26.2\%}    & \redtext{26.3\%}    \\ \hline
Tenset & 17.6\% & 25.8\% & 22.1\% & 23.8\% & 10.1\% & 30.5\%  & \redtext{10.1\%}  & 32.3\%  & 30.4\%    & 42.2\%    \\ \hline
TLP    & 10.1\% & 22.3\% & \redtext{9.1\%}  & 16.5\% & 18.0\% & 26.9\%  & 19.4\%  & \redtext{10.0\%}  & 41.7\%    & 30.0\%    \\ \hline
\end{tabular}%
}
\end{table}

\section{Conclusion}\label{sec:conclusion}
In this work, we present LLMulator, a novel framework that redefines the paradigm of performance prediction for dataflow accelerators through progressive numeric modeling. By synergizing the semantic understanding capabilities of pre-trained language models with domain-specific innovations, our approach addresses three fundamental challenges in cost modeling: application generalization, input-aware dynamic calibration, and generalization of hardware architecture and mapping parameters. Extensive experiments across ASIC design spaces demonstrate that LLMulator achieves state-of-the-art prediction error with 12.2\% MAPE, outperforming existing methods (TLP, GNNHLS) by 16.7\% and 7.8\%.

\begin{acks} 
We would like to acknowledge the financial support from the NSFC (Grant No. 62222411), the National Key R\&D Program of China (Project No. 2023YFB4404400), and the reviewers of MICRO 2025 for their valuable feedback.
\end{acks}

\bibliographystyle{ACM-Reference-Format}
\bibliography{sample-base}

\end{document}